\documentclass[useAMS,usenatbib]{mn2e}

\usepackage{amsmath, amssymb}
\usepackage{changepage}
\usepackage{graphicx}
\usepackage{aas_macros}

\newcommand{\Rp}{R_p}
\newcommand{\Rs}{R_{\star}}
\newcommand{\RpRs}{\Rp/\Rs}
\newcommand{\Fp}{F_p}
\newcommand{\Fs}{F_{\star}}
\newcommand{\FpFs}{\Fp/\Fs}
\newcommand{\aRs}{a/\Rs}
\newcommand{\Tmid}{T_{\textnormal{mid}}}
\newcommand{\um}{\mu\textnormal{m}}
\newcommand{\tn}{\textnormal}
\newcommand{\bigO}{\mathcal{O}} % symbol for big 'O'
\newcommand{\alphab}{\boldsymbol{\alpha}} % symbol for alpha vector
\newcommand{\gammab}{\boldsymbol{\gamma}} % symbol for gamma vector
\newcommand{\vi}{\boldsymbol{v}_{\boldsymbol{i}}} % symbol for v_i vector
\newcommand{\vj}{\boldsymbol{v}_{\boldsymbol{j}}} % symbol for v_j vector

\newcommand{\mb}{\mathbf}
\newcommand{\gb}{\boldsymbol}

\title[A uniform analysis of HD\,209458b \textit{Spitzer}/IRAC lightcurves with Gaussian process models]{A uniform analysis of HD\,209458b \textit{Spitzer}/IRAC lightcurves with Gaussian process models}
\author[Evans et al.]{ Thomas M. Evans,$^{1,2}$\thanks{E-mail:
tevans@astro.ex.ac.uk} Suzanne Aigrain,$^{1}$ Neale Gibson,$^{3}$ Joanna K. Barstow,$^{1}$ \newauthor David S.\ Amundsen,$^{2}$ Pascal Tremblin$^{2,4}$ and Pierre Mourier$^{2,5}$ \\
$^{1}$Department of Physics, University of Oxford, Denys Wilkinson Building, Keble Road, Oxford OX1 3RH, UK\\
$^{2}$School of Physics, University of Exeter, EX4 4QL Exeter, UK\\
$^{3}$European Southern Observatory, Karl-Schwarzschild-Strasse 2, D-85748 Garching, Germany\\
$^{4}$Maison de la Simulation, CEA-CNRS-INRIA-UPS-UVSQ, USR 3441, Centre d'\'{e}tude de Saclay, 91191 Gif-Sur-Yvette, France\\
$^{5}$Master ICFP, D\'{e}partement de Physique, Ecole Normale Sup\'{e}rieure, 24 Rue Lhomond, 75005 Paris, France}

\begin{document}

\date{Accepted, 21 April 2015. Received, 17 April 2015; in original form, 16 March 2015}

\pagerange{\pageref{firstpage}--\pageref{lastpage}} \pubyear{2015}

\maketitle

\label{firstpage}

\begin{abstract}
We present an analysis of \textit{Spitzer}/IRAC primary transit and secondary eclipse lightcurves measured for HD\,209458b, using Gaussian process models to marginalise over the intrapixel sensitivity variations in the $3.6\um$ and $4.5\um$ channels and the ramp effect in the $5.8\um$ and $8.0\um$ channels. The main advantage of this approach is that we can account for a broad range of degeneracies between the planet signal and systematics without actually having to specify a deterministic functional form for the latter. Our results do not confirm a previous claim of water absorption in transmission. Instead, our results are more consistent with a featureless transmission spectrum, possibly due to a cloud deck obscuring molecular absorption bands. For the emission data, our values are not consistent with the thermal inversion in the dayside atmosphere that was originally inferred from these data. Instead, we agree with another re-analysis of these same data, which concluded a non-inverted atmosphere provides a better fit. We find that a solar-abundance clear-atmosphere model without a thermal inversion underpredicts the measured emission in the $4.5\um$ channel, which may suggest the atmosphere is depleted in carbon monoxide. An acceptable fit to the emission data can be achieved by assuming that the planet radiates as an isothermal blackbody with a temperature of $1484\pm 18$\,K.
\end{abstract}

\begin{keywords}
planets and satellites:\ atmospheres; planets and satellites:\ general; methods:\ data analysis, astronomical instrumentation, methods, and techniques; stars:\ individual:\ HD209458
\end{keywords}

\section{Introduction} \label{sec:intro}

Over the past decade, the \textit{Spitzer Space Telescope} has proven to be a productive facility for characterising the atmospheres of transiting exoplanets \citep[e.g.][]{2005ApJ...626..523C, 2006ApJ...644..560D, 2007Natur.447..183K, 2009ApJ...699..478D, 2012ApJ...752...81C, 2013ApJ...766...95L, 2014ApJ...796..100T}. The ability of its instruments to probe the $\sim$3--25$\um$ wavelength range has provided constraints on the thermal emission of numerous exoplanets, as well as atmospheric transmission in a region dominated by absorption from molecular species such as water, methane, carbon monoxide, and carbon dioxide. Hot Jupiters have offered especially favourable targets for such observations, given their large atmospheric scale heights and relatively strong emission at these wavelengths.

This paper focuses on observations made with the Infrared Array Camera (IRAC), which has been the most widely used \textit{Spitzer} instrument for observing exoplanets. Specifically, we analyse ten transits and eleven eclipses that have been measured for HD\,209458b, most of which have already been published \citep{2008ApJ...673..526K, 2010MNRAS.409..963B, 2014ApJ...790...53Z, 2014ApJ...796...66D}. By providing a uniform analysis of these datasets, we aim to re-evaluate a number of claims that have been made in the literature. In particular, \cite{2010MNRAS.409..963B} measured significantly larger effective radii for the planet in the $5.8\um$ and $8.0\um$ channels relative to the $3.6\um$ and $4.5\um$ channels in transmission, and interpreted this as evidence for water absorption. However, \cite{2013ApJ...774...95D} have since resolved the water absorption band centred at $1.4\um$ using the \textit{Hubble Space Telescope} (HST) Wide Field Camera 3 (WFC3), and found that it has a much lower amplitude than would be expected based on the results of Beaulieu et al. We also seek to address the claim of a thermal inversion in the dayside atmosphere, which was first postulated by \cite{2008ApJ...673..526K} based on the deeper eclipses those authors measured for the $4.5\um$ and $5.8\um$ channels relative to the $3.6\um$ channel. The former channels coincide with absorption features due to water and carbon monoxide:\ therefore, seeing these features in emission would suggest an increasing temperature profile with decreasing pressure. \cite{2014ApJ...796...66D} have challenged this picture by presenting revised eclipse depths that are suggestive of a non-inverted pressure-temperature profile. Furthermore, \cite{2014MNRAS.444.3632H} have suggested that the emission data for HD\,209458b are consistent with radiation from an isothermal blackbody.

The conflicting results obtained by different authors analysing the same datasets is likely due to the various methods that have been used to account for the instrumental systematics that dominate IRAC lightcurves. The main contribution of the current study is to apply the machinery of Gaussian processes (GPs) to the task of treating these systematics. This work follows similar applications of GP models to transit lightcurves published by \cite{2012MNRAS.419.2683G, 2012MNRAS.422..753G, 2013MNRAS.428.3680G, 2013MNRAS.436.2974G}, \cite{2013ApJ...772L..16E}, and \cite{2014MNRAS.445.3401G}. One of the primary advantages of GP models is that they allow us to naturally handle correlations in the data that may be poorly understood from a first principles standpoint, by specifying only high-level properties of the covariance. This relaxes the assumptions built into our model, by removing the need to associate the systematics with a deterministic functional form. The resulting model is less restrictive, allowing us to capture a broad range of systematics behaviours with a relatively small number of tunable parameters. Furthermore, GPs are Bayesian models in the sense that uncertainty is treated transparently using self-consistent rules of probability. Each unknown in our model is associated with a probability distribution that reflects our uncertainty in its value, and it is possible to write down an expression for the likelihood of the observed data given specific values for the model parameters, i.e.\ the model posterior distribution. We can then optimise the model posterior with respect to the unknown parameters, or marginalise over the parameter space using a method such as Markov chain Monte Carlo (MCMC). These properties make GP models suitable for inferring planet parameters from transit lightcurves affected by systematics that are not especially well-understood, such as those obtained with IRAC. 

The paper is arranged as follows. Section \ref{sec:observations} describes the lightcurve observations analysed for this study and Section \ref{sec:datared} describes how we produced lightcurves from the raw data frames. IRAC instrumental systematics are described in Section \ref{sec:systematics}, with an overview of methods that have been used to correct for them previously in the literature. Section \ref{sec:gpmodels} outlines the GP methodology that we adopt in the current study, and Section \ref{sec:lcfitting} describes the lightcurve fitting. The results are presented in Section \ref{sec:results} and discussed in Section \ref{sec:discussion}, with a focus on the implications for the planet atmosphere. Our conclusions are summarised in Section \ref{sec:conclusion}.

\section{Observations} \label{sec:observations}

\begin{table*}
\begin{minipage}{\linewidth} 
\centering
\caption{HD209458 datasets analysed for this study. \label{table:irac_datasets}}
\begin{tabular}{cccccccc}
\\
\hline \\
                          &     &         & \multicolumn{4}{c}{Channels ($\um$)} & \medskip \\ \cline{4-7} 
\\
Program & P.I. & Type & 3.6 & 4.5 & 5.8 & 8.0 & References\footnote{Be10 \citep{2010MNRAS.409..963B}, DL14 \citep{2014ApJ...796...66D}, Kn08 \citep{2008ApJ...673..526K}, Ze14 \citep{2014ApJ...790...53Z}.} \smallskip \\
\hline \\
20523  & D.~Charbonneau & Eclipse & Yes & Yes & Yes & Yes &  Kn08, DL14 \\
                        40280 & H.~Knutson         & Half-phase & - & - & - & Yes &  DL14  \\ 
                        461     & G.~Tinetti            & Transit & Yes & Yes & Yes & Yes &  Be10  \\ 
                        60021 & H.~Knutson         & Full-phase & Yes & Yes & - & - &  Ze14, DL14 \\ \\ \hline
\end{tabular}
\end{minipage}
\end{table*}

We have analysed ten primary transits and eleven secondary eclipses for HD209458b, made across all four IRAC channels. Details of the relevant \textit{Spitzer} observing programs are given in Table \ref{table:irac_datasets}, along with references to previously published analyses. More specific information for the individual lightcurves is given in Table \ref{table:irac_lightcurves}. For this study, we did not model the complete half- and full-phase datasets acquired for Programs 30825, 40280, and 60021. Instead, for these lightcurves only $\sim$5\,hr subsections centred on the transits and eclipses were analysed.

Most observations were made in stare mode. The only exceptions were those made as part of Program 20523, which have been published in \cite{2008ApJ...673..526K}. For the latter, four sets of 64 frames were acquired in a given channel before the telescope was repointed to be centred on the next channel, and another four sets of 64 frames were taken. This process was cycled through each of the four channels, and repeated for the duration of the observations. Knutson et al.\ discarded the first set of 64 frames in each set of four, as the star was still drifting significantly during this time following the repointing. In addition, those authors discarded the first ten frames and the 58th frame from each set of 64 for the 5.8$\um$ and 8.0$\um$ channels, as these exhibited count levels consistently below the median. We performed two separate analyses for these lightcurves:\ one with this culling applied, and one without. However, we obtained consistent results in both cases, and only present results for the unculled dataset below. 

\begin{table*} 
\begin{minipage}{\linewidth} 
\centering
\caption{Lightcurve details. \label{table:irac_lightcurves}}
\begin{tabular}{cccccccccc}
\footnotesize
\\

\hline \\
\rule{0pt}{4ex} Program & Type & Channel & Date & Mode\footnote{Readout mode and frame time in seconds.} & Flagged\footnote{Fraction of frames flagged as bad.} & $N_b$\footnote{Number of frames in the binned dataset used for the GP lightcurve fit.} & $\eta_b$\footnote{Number of consecutive frames per bin.} & $\Delta t_{\tn{med}}$\footnote{Median cadence of binned frames.} & $r_{\tn{ap}}$\footnote{Photometric aperture radius.}\\
 & & ($\um$) & (UT) & &(\%) &&& (sec) & (pix) \smallskip \\
\hline \\
20523  & Eclipse       & 3.6  & 2005 Nov 28 & sub, 0.1   & 0.36 & 1115 & 32 & 4.5  & 4.0  \\ 
       & Eclipse       & 4.5  & 2005 Nov 28 & sub, 0.1   & 0.21 & 1115 & 32 & 4.5  & 3.0 \\ 
       & Eclipse       & 5.8  & 2005 Nov 28 & sub, 0.1   & 0.32 & 1115 & 32 & 4.5  & 2.5 \\ 
       & Eclipse       & 8.0  & 2005 Nov 28 & sub, 0.1   & 0.55 & 1115 & 32 & 4.5  & 3.0 \smallskip\\ 
40280  & Transit       & 8.0  & 2007 Dec 25 & sub, 0.4   & 1.11 & 1577 & 32 & 13.5 & 3.5   \\
       & Eclipse       & 8.0  & 2007 Dec 24 & sub, 0.4   & 1.12 & 1577 & 32 & 13.5 & 4.0   \smallskip\\ 
461    & Transit       & 3.6  & 2007 Dec 31 & full, 0.4  & 1.68 & 1427 & 2 & 16.8 & 2.5   \\ 
       & Transit       & 3.6  & 2008 Jul 19 & full, 0.4  & 2.03 & 1428 & 2 & 16.8 & 2.5    \\ 
       & Transit       & 4.5  & 2008 Jul 22 & full, 0.4  & 1.67 & 1288 & 2 & 16.8 & 2.5    \\
       & Transit       & 5.8  & 2007 Dec 31 & full, 2.0  & 3.78 & 1425 & 2 & 16.8 & 3.5    \\
       & Transit       & 5.8  & 2008 Jul 19 & full, 2.0  & 3.81 & 1425 & 2 & 16.8 & 3.0    \\ 
       & Transit       & 8.0  & 2008 Jul 22 & full, 2.0  & 4.42 & 1286 & 2 & 16.8 & 4.0   \smallskip\\ 
60021  & Eclipse (1st) & 3.6  & 2011 Jan 12 & sub, 0.1   & 0.25 & 1285 & 128 & 16.8 & 2.5  \\
       & Transit       & 3.6  & 2011 Jan 14 & sub, 0.1   & 0.25 & 1286 & 128 & 16.8 & 2.5  \\
       & Eclipse (2nd) & 3.6  & 2011 Jan 16 & sub, 0.1   & 0.29 & 1286 & 128 & 16.8 & 3.0  \\
       & Eclipse (1st) & 3.6  & 2014 Feb 13 & sub, 0.1   & 0.23 & 1286 & 128 & 16.8 & 3.0 \\ 
       & Transit       & 3.6  & 2014 Feb 15 & sub, 0.1   & 0.23 & 1286 & 128 & 16.8 & 3.0  \\
       & Eclipse (2nd) & 3.6  & 2014 Feb 17 & sub, 0.1   & 0.18 & 1285 & 128 & 16.8 & 2.5  \\
       & Eclipse (1st) & 4.5  & 2010 Jan 18 & sub, 0.4   & 0.62 & 1577 & 32 & 13.6 & 2.5   \\ 
       & Transit       & 4.5  & 2010 Jan 19 & sub, 0.4   & 0.73 & 1575 & 32 & 13.6 & 3.0   \\ 
       & Eclipse (2nd) & 4.5  & 2010 Jan 21 & sub, 0.4   & 0.62 & 1577 & 32 & 13.6 & 2.5   \\ \\ \hline 
\end{tabular}
\end{minipage}
\end{table*}

\section{Data reduction} \label{sec:datared}

The Basic Calibrated Data (BCD) frames for each lightcurve were reduced using a custom pipeline written in the Python programming language.\footnote{Publicly available at www.github.com/tomevans} The first step performed by the pipeline is to calculate the background level and locate the stellar centroid in each BCD frame. The background was estimated by taking the median pixel value from the four $8 \times 8$\,pixel subarrays at the corners of each frame, and then subtracted from each pixel in the array. The centroid coordinates were then determined by taking the flux-weighted mean of a $7 \times 7$\,pixel subarray centred on the approximate location of the star. An initial guess was provided for the stellar centroid coordinates in the first frame, and coordinates determined for the previous frame were used as the initial guess in subsequent frames. Mid-times were computed for each exposure in Barycentric Julian Date Coordinated Universal Time (BJD$_{\textnormal{UTC}}$) using the BMJD\_OBS and FRAMTIME header entries. Bad frames were flagged by identifiying those with outlying centroid coordinates or pixel counts. This was done by comparing against the median and standard deviation of the 30 frames immediately preceding and following each frame. If the centroid coordinates or any pixel counts within a subarray spanning the photometric aperture centred at the stellar centroid differed from the median by $>5\sigma$, the frame was discarded from the analysis. This process was iterated twice, resulting in 0.2--4.4\% of the frames being discarded depending on the dataset (Table \ref{table:irac_lightcurves}).

\begin{figure*}
\centering  % this centres figure in column
\includegraphics[width=\linewidth]{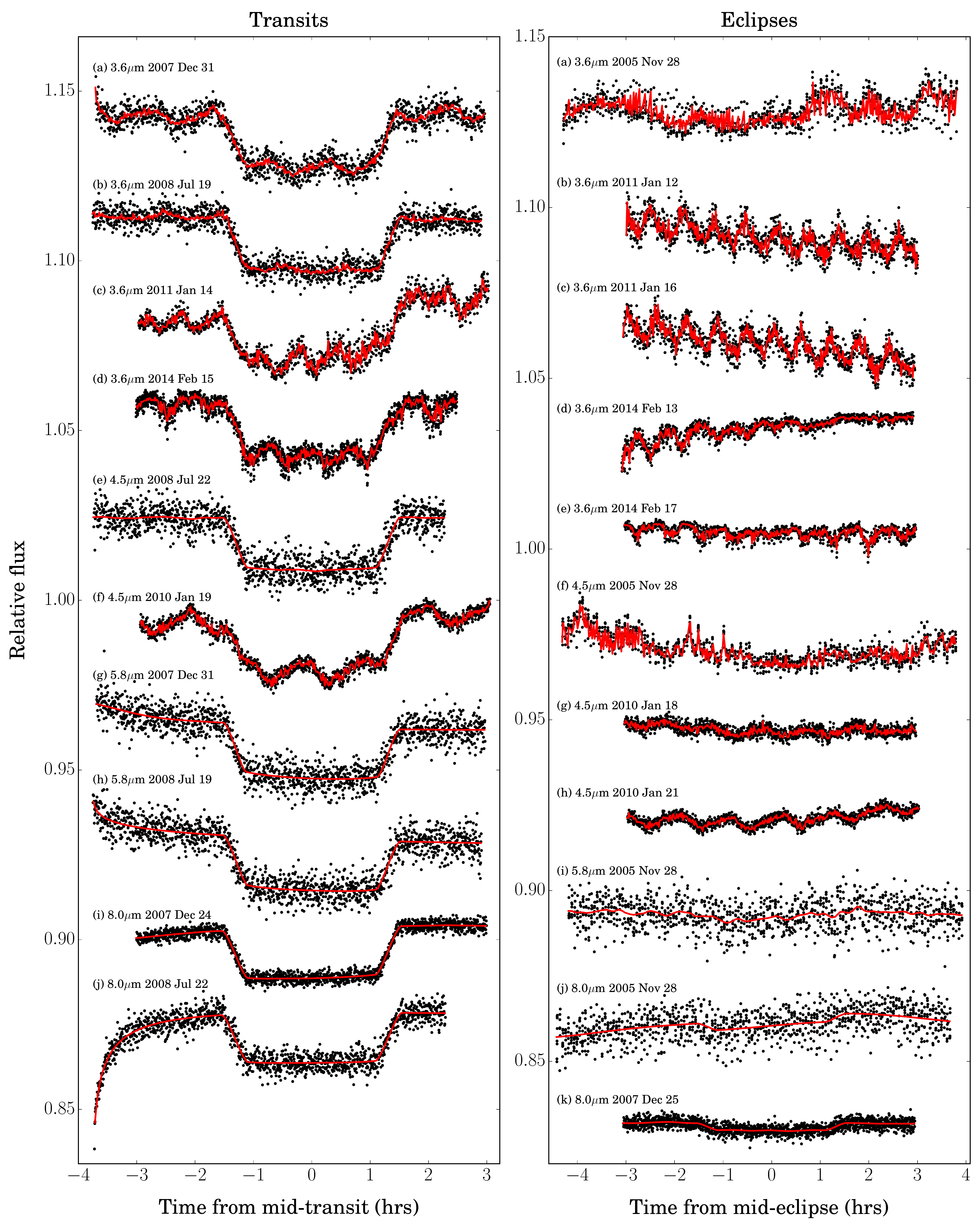}
\caption{Raw lightcurves obtained for the 10 transits (left column) and 11 eclipses (right column) analysed in this study. Red lines show best-fit GP models, which are described in Section \ref{sec:gpmodels}. The lightcurves shown in this plot have been binned in time, which was done to make the GP model fitting computationally tractable (Section \ref{sec:lcbinning}). The plotted GP models have been further binned into 1 minute bins and vertical offsets have been applied to each lightcurve for clarity.}
\label{fig:lcs_raw}
\end{figure*}

Photometry was performed for each remaining frame by summing the pixel counts within circular apertures. Separate reductions were obtained for different aperture radii, ranging between 2--6\,pixel in increments of 0.5\,pixel. Due to the undersampled nature of the IRAC point spread function (PSF), we linearly interpolated the native pixel array onto a $10 \times 10$ super-sampled grid, as has been done by others previously \citep[e.g.][]{2010Natur.464.1161S}. These interpolated sub-pixels were counted towards the aperture sum if their centres fell within the aperture radius. The resulting lightcurves are shown in Figure \ref{fig:lcs_raw}.

\section{Instrumental systematics} \label{sec:systematics}

The raw lightcurves are affected by instrumental systematics that are characteristic of IRAC and have been documented extensively in the literature \citep[e.g.][]{2005ApJ...626..523C, 2010ApJ...721.1861A, 2010ARA&A..48..631S, 2012ApJ...754..136S}. The systematics divide into two broad categories:\ intrapixel sensitivity variations in the $3.6\um$ and $4.5\um$ channels, and the ramp effect in the $5.8\um$ and $8.0\um$ channels.

\subsection{$3.6\um$ and $4.5\um$ channels} \label{sec:ch36_ch45_systematics}

In the $3.6\um$ and $4.5\um$ channels, which employ InSb detectors, the measured flux correlates with the position of the stellar PSF on the detector array. As noted in the IRAC instrument handbook,\footnote{http://irsa.ipac.caltech.edu/data/SPITZER/docs/irac} this effect is believed to be caused by variations in the quantum efficiency across individual pixels. Pointing drift during observations, combined with the undersampled PSF, therefore results in variations in the measured flux at the level of a few percent. 

Traditionally, intrapixel sensitivity variations have been treated in IRAC data by decorrelating the lightcurve against a low-order polynomial in the $xy$ centroid coordinates of the stellar PSF, which can either be removed prior to fitting the planet signal or fit simultaneously with the planet signal \citep[e.g.][]{2008ApJ...686.1341C, 2008ApJ...673..526K, 2009ApJ...699..478D}. The main issue with this method is that a low-order polynomial may not have the flexibility to fully capture the underlying correlations in the data if there is fine-scale structure present. This could be addressed to some extent by continuing to add higher order terms to the polynomial decorrelation; however, such an approach runs the risk of overfitting and increases the dimensionality of the parameter space that must be marginalised over using a method such as MCMC. 

An alternative decorrelation method has been suggested by \cite{2010PASP..122.1341B}, which uses a 2D Gaussian convolution to construct a smoothed pixel sensitivity map from the flux measurements. Using this approach, the authors identified a high-frequency corrugation structure in the $xy$ sensitivity for a $4.5\um$ lightcurve that would not have been captured by a low-order polynomial decorrelation. Along similar lines, \cite{2012ApJ...754..136S} have proposed the use of bilinear interpolation of the measured fluxes to map intrapixel sensitivity variations, which is sensitive to spatial scales corresponding to the grid of knots used. \cite{2012ApJ...754...22Kmnras} and \cite{2013ApJ...766...95L} have also identified a correlation between the noise pixel parameter $\beta$ of a given data frame -- which is inversely proportional to the PSF sharpness parameter, described in \cite{1974JOSA...64.1200M} -- and the measured flux. This can be treated by either using variable photometric apertures for each frame that scale with $\beta$, or including $\beta$ as a decorrelation variable in the systematics treatment. More recently, \cite{2014arXiv1411.7404D} have presented a pixel-level decorrelation (PLD) method, which works by linearly decorrelating the measured fluxes against the individual pixel counts within a subarray centred on the PSF, rather than using the centroid coordinates directly.

\subsection{$5.8\um$ and $8.0\um$ channels} \label{sec:ch58_ch80_systematics}

In the $5.8\um$ and $8.0\um$ channels, which employ Si:As detectors, the measured flux smoothly increases or decreases before levelling off. As with the intrapixel sensitivity variations, the amplitude of this effect is usually a few percent, with the steepest change in flux occurring during the initial $\sim 1$\,hr of observations. The ramp can be attributed to the sensitivity of individual pixels varying as a function of time, where the rate of change depends on the illumination level of the pixel \citep[e.g.][]{2007Natur.447..183K}. 

It has been suggested that the ramp is caused by electrons getting caught in charge traps within the detector \citep[e.g.][]{2007ApJ...667L.199D, 2010ApJ...721.1861A, 2010ARA&A..48..631S}. As more photons arrive, the charge traps fill up, resulting in less electrons getting trapped and a higher flux being measured. However, \cite{2010ARA&A..48..631S} acknowledge that the physics of the detectors are poorly understood and this simple picture may be incomplete or wrong. For instance, lightcurves measured in the 5.8$\um$ channel exhibit a ramp-down behaviour (Figure \ref{fig:lcs_raw}), which is not obviously explained by charge trapping.

Standard practise has been to fit $5.8\um$ and $8.0\um$ lightcurves by multiplying the transit signal by a parametric model that provides a good approximation to the ramp. One approach is to remove the initial steep section of the lightcurve and fit the remainder with a linear or quadratic polynomial in time \citep[e.g.][]{2007ApJ...667L.199D, 2010MNRAS.409..963B}. However, a more common approach is to model the full baseline with a low-order polynomial in logarithmic time \citep[e.g.][]{2008ApJ...686.1341C, 2008ApJ...673..526K, 2009ApJ...699..478D, 2010ApJ...711..111M} or exponential time \citep[e.g.][]{2010ApJ...721.1861A}.

Subtle differences between the various ramp functions used to model $5.8\um$ and $8.0\um$ lightcurves can potentially bias the inferred planet parameters. For example, polynomials in exponential time tend to retain slightly more curvature than polynomials in logarithmic time after the initial steep gradient. This can result in an underestimated transit depth if the ramp is decreasing, or vice versa if the ramp is increasing. To try avoid effects such as these, the lightcurve analysis is often performed separately for a number of different ramp functions, and the one that minimises the model residuals or maximises an approximation to the Bayesian model evidence, such as the negative Bayesian information criterion \citep{Schwarz1978}, is retained \citep[e.g.][]{2012ApJ...754..136S}. 

\section{Gaussian processes for IRAC systematics} \label{sec:gpmodels}

In this paper, we present an alternative method for treating intrapixel sensitivity variations and the ramp effect in IRAC lightcurves based on Gaussian processes (GPs). The first application of GP models to transit lightcurves was made by \cite{2012MNRAS.419.2683G}, with subsequent applications in \cite{2012MNRAS.422..753G, 2013MNRAS.428.3680G, 2013MNRAS.436.2974G}, \cite{2013ApJ...772L..16E}, and \cite{2014MNRAS.445.3401G}, to which the reader is referred for further details.

Formally, a GP is defined as a collection of data points, any subset of which has a multivariate normal distribution \citep[e.g.][]{gpml06}. Indeed, this assumption is regularly made for transit lightcurve analyses published in the literature, even if it is not stated explicitly. It is equivalent to assuming that what we measure is some underlying signal, which is the combination of the astrophysical signal of interest and additional systematic terms, plus uncorrelated Gaussian noise. If we are in possession of a model $\mu$ that describes the underlying signal, the probability of measuring a specific dataset $\mb{d}= \{ d_1, d_2, \ldots, d_N \} $ with Gaussian errorbars $\gb{\sigma_w} = \{ \sigma_{w,1}, \sigma_{w,2}, \ldots, \sigma_{w,N} \}$ for a set of model parameters $\gb{\alpha} = \{ \alpha_1, \alpha_2, \ldots, \alpha_M \}$ is given by:
\begin{eqnarray}
 p ( \mb{d} | \gb{\alpha} ) \ = \ \prod\limits_{i=1}^N{\frac{1}{\sigma_{w,i}\sqrt{2\pi}}}\exp\left[-\frac{(d_i-\mu_i)^2}{2\sigma_{w,i}^2}\right] \ = \ \mathcal{N}\left( \gb{\mu}, \gb{\Sigma} \right) \ , &&
\end{eqnarray}
where $\mathcal{N}$ denotes a multivariate normal distribution and $\gb{\Sigma}=\tn{diag}[\gb{\sigma_w}]$ is the diagonal covariance matrix. For fixed errorbars, optimising the log likelihood is therefore identical to the familiar practice of minimising the $\chi^2$ statistic, since:
\begin{eqnarray}
\ln \mathcal{N}\left( \gb{\mu}, \gb{\Sigma} \right) & = & -\frac{1}{2}\chi^2 - \frac{1}{2}\sum\limits_{i=1}^N{\ln\sigma_{w,i}} - \frac{1}{2}N\ln 2\pi \ . \label{eq:logp_gp_diagonal}
\end{eqnarray}
Note, however, that Equation \ref{eq:logp_gp_diagonal} requires us to build all of our information about the underlying signal into the deterministic mean function $\gb{\mu}$. This is not desirable if, as is commonly the case, the systematics are poorly understood from first principles and an explicit functional form is not available to describe them.

An alternative option is to incorporate the systematics into our model by allowing for nonzero off-diagonal entries in the covariance matrix of the likelihood function, such that:
\begin{eqnarray}
\ln p ( \mb{d} | \gb{\alpha}, \gb{\gamma} ) & = & \ln \mathcal{N}\left( \gb{\mu}, \mb{K} + \gb{\Sigma} \right) \ , \label{eq:logp_gp_general}
\end{eqnarray}
where $K_{ij}$ gives the covariance between the $i$th and $j$th data points, and $\gb{\gamma}$ are the parameters that control the behaviour of the covariance. Although we no longer have to provide an explicit functional form for the systematics contribution, we must now specifiy a kernel function to populate the entries of the covariance matrix $\mb{K}$. However, by modelling the statistical covariance between data points rather than the deterministic systematics signal directly, it is possible to capture a broad range of behaviours with relatively few free parameters. Thus, GP models are simultaneously parsimonious and flexible.

Before proceeding to describe the specific mean functions and covariance kernels adopted in the current study in Sections \ref{sec:meanfunctions} and \ref{sec:covkernels}, it is worth also pointing out the fact that GPs have the desirable property of automatically implementing the principle of Occam's razor. This can be seen if we expand Equation \ref{eq:logp_gp_general} into its constituent terms:
\begin{eqnarray}
\ln p ( \mb{d} | \gb{\alpha}, \gb{\gamma} )=-\frac{1}{2}\mb{r}^{\tn{T}}\left( \mb{K} + \gb{\Sigma} \right)^{-1} \mb{r} - \frac{1}{2}\ln | \mb{K}+\gb{\Sigma} | - \frac{1}{2} N \ln 2 \pi  \ , && \label{eq:logp_gp_general_expanded}
\end{eqnarray}
where $\mb{r}$ is a vector containing the model residuals, with $i$th term given by $r_i = d_i-\mu_i$. The first term on the righthand side, $-\frac{1}{2}\mb{r}^{\tn{T}}(\mb{K}+\gb{\Sigma})^{-1}\mb{r}$, serves as a goodness-of-fit term. For a given covariance matrix, it increases as the residuals become smaller, rewarding mean functions $\gb{\mu}$ that match the data well. The second term, $-\frac{1}{2}\ln|\mb{K}+\gb{\Sigma}|$, can be thought of as a complexity penalty. This is because increasing the complexity of a model is equivalent to assigning similar probabilities to an increasing diversity of functions. In other words, as the model complexity increases, the likelihood function becomes less sharply peaked near the mean $\gb{\mu}$ and the probability mass of the model becomes more diffusely spread throughout the function space. This is precisely what happens as the term $-\frac{1}{2}\ln|\mb{K}+\gb{\Sigma}|$ decreases, in effect penalising model complexity. Finally, the third term, $-\frac{1}{2}N\ln 2\pi$, remains constant for a given dataset. The balance between the first two terms of Equation \ref{eq:logp_gp_general_expanded} therefore ensures that the probability mass of the model is distributed over the parameter space in a manner that optimises the trade-off between goodness-of-fit and model complexity.

\subsection{Mean functions} \label{sec:meanfunctions}

The mean function $\gb{\mu}$ defines the model for the astrophysical signal with well-understood form; namely, a primary transit or secondary eclipse. For this purpose, we adopt the analytic transit functions of \cite{2002ApJ...580L.171M}. We set the orbital eccentricity to zero, consistent with observational evidence \citep[e.g.][]{2011MNRAS.414.1278P}. We fix the orbital period to $P=3.52474859$\,day \citep{2007ApJ...655..564K}. 

For the primary transits, the mean function parameters that are allowed to vary are the radius ratio $\RpRs$, normalised semimajor axis $\aRs$, impact parameter $b = a\cos i/\Rs$, and transit mid-time $\Tmid$, such that $\gb{\alpha} = \{ \RpRs, \aRs, b, \Tmid \}$ in Equation \ref{eq:logp_gp_general_expanded}. Stellar limb darkening is treated using the nonlinear law of \cite{2004A&A...428.1001C}, with coefficients fixed to those provided by \cite{2012A&A...539A.102H} which were obtained specifically for the IRAC bandpasses using a 3D stellar model for HD\,209458 (Table \ref{table:ld_coeffs}).

\begin{table}
\begin{minipage}{\columnwidth} 
\centering
\caption{Nonlinear limb darkening coefficients. \label{table:ld_coeffs}}
\begin{tabular}{ccccc} 
%\scriptsize \\
\hline \\
 & \multicolumn{4}{c}{Channels ($\um$)} \medskip \\ \cline{2-5} 
\\
& 3.6 & 4.5 & 5.8 & 8.0 \smallskip \\
\hline \\
$c_1$ &  $0.5564$ &  $0.4614$ &  $0.4531$ &  $0.4354$ \\
$c_2$ & $-0.5462$ & $-0.4277$ & $-0.5119$ & $-0.6067$ \\
$c_3$ &  $0.4315$ &  $0.3362$ &  $0.4335$ &  $0.5421$ \\
$c_4$ & $-0.1368$ & $-0.1074$ & $-0.1431$ & $-0.1816$ \\ \\ \hline
\end{tabular}
\end{minipage}
\end{table}

For the secondary eclipses, only the eclipse depth $\FpFs$ and eclipse mid-time $\Tmid$ are allowed to vary, such that $\gb{\alpha} = \{ \FpFs, \Tmid  \}$. The remaining mean function parameters are fixed to the values published by \cite{2008ApJ...677.1324T}, namely, $\RpRs=0.121$, $\aRs = 8.76$ and $b=0.507$.

\subsection{Covariance kernels} \label{sec:covkernels}

Entries of the covariance matrix $\mb{K}$ are constructed using a kernel function, such that $K_{ij} = k(\mb{v_i},\mb{v_j})$, where $\mb{v_i}$ and $\mb{v_j}$ are vectors of inputs associated with the $i$th and $j$th data points, respectively. By inputs, we refer to variables that correlate with the measured signal -- these are typically the same variables that would be used for standard polynomial systematics decorrelations. Further discussion of common GP kernels, such as the squared exponential and Mat\'{e}rn kernels, can be found in \cite{2012MNRAS.419.2683G}. 

In this study, we parameterise the entries of the covariance matrix $\mb{K}$ for the $3.6\um$ and $4.5\um$ channels as the sum of two kernels: a squared exponential kernel with the centroid $xy$ coordinates as inputs and a Mat\'{e}rn $\nu = 3/2$ kernel with time $t$ as the input. Writing this out explicitly, the combined kernel is given by:
\begin{eqnarray}
k\left( \, \vi \, , \, \vj \, \right) &=& k_{xy} + k_{t} \ , \label{eq:ch36_ch45_kernel}
\end{eqnarray}
where:
\begin{eqnarray}
k_{xy} &=&   A_{xy}^2\,\exp\left[ \, - \left( \frac{x_i-x_j}{L_x} \right)^2 - \left( \frac{y_i-y_j}{L_y} \right)^2 \, \right]  \ , \label{eq:kxy} \\
&& \nonumber \\
k_{t} &=& A_t^2\,\left[ \, 1\,+\,\frac{t_i-t_j}{L_t}\,\sqrt{3} \, \right]\,\exp\left[  \, -\left( \frac{t_i-t_j}{L_t}\right) \, \sqrt{3} \,  \right] \ , \label{eq:kt}
\end{eqnarray}
such that $\gammab = \{ \, A_{xy} \, , \, L_x \, , \, L_y \, , \, A_t \, ,  \, L_t \, \}$ in the notation of Equation \ref{eq:logp_gp_general_expanded}. The squared exponential component of this kernel accounts for the smooth spatial variations in pixel sensitivities that dominate the systematics of the $3.6\,\um$ and $4.5\,\um$ channels, while the Mat\'{e}rn component accounts for any residual correlated noise in the lightcurve. 

For the 5.8$\um$ and 8.0$\um$ channel lightcurves, we used a squared exponential kernel to model the dominant ramp effect, with form given by:
\begin{eqnarray}
k_{\tau} &=&   A_{\tau}^2\,\exp\left[ \, -\left( \frac{\tau_i-\tau_j}{L_\tau} \right)^2\, \right]  \ , \label{eq:ktau} 
\end{eqnarray}
where $\tau=\ln \left( t+h \right)$ is logarithmic time $t$, and $h$ is a parameter that can be inferred from the data (see below). Parameterising the covariance with $k_\tau$ allows us to capture the dominant behaviour of the ramp effect; namely, a steep initial gradient followed by a levelling off of the measured flux (Figure \ref{fig:lcs_raw}). However, we stress that by parameterising the covariance between data points according to Equation \ref{eq:ktau}, we are not constraining the systematics to be monotonically increasing or decreasing in time. This is a valuable property of GPs given that the ramp effect is not necessarily strictly monotonic, with an overshoot effect identified in a number of datasets \citep[e.g.][]{2012ApJ...754...22Kmnras}. As with the other channels, we also include a time-dependent Mat\'{e}rn $\nu = 3/2$ kernel $k_t$ to account for residual correlations in the lightcurve that may not be related to the ramp. Therefore, the final kernel is given by:
\begin{eqnarray}
k\left( \, \vi \, , \, \vj \, \right) &=& k_{\tau} + k_{t} \ , \label{eq:ch58_ch80_kernel}
\end{eqnarray}
with covariance parameters $\gammab = \{ \, A_{\tau} \, , \, L_\tau \, , \, h \, , \, A_t \, ,  \, L_t \, \}$.

The covariance kernels outlined above allow for systematics treatments that are at least as versatile as others used in the literature. For instance, the $k_{xy}$ component of Equation \ref{eq:ch36_ch45_kernel} is similar in concept to the 2D Gaussian correction developed by Ballard et al.\ (Section \ref{sec:ch36_ch45_systematics}). Similarly, the $k_{\tau}$ component of Equation \ref{eq:ch58_ch80_kernel} is reminiscent of the logarithmic time polynomials used in other published studies (Section \ref{sec:ch58_ch80_systematics}). However, as it is only the covariance between data points that is parameterised in terms of logarithmic time, the underlying signal itself need not be monotonically increasing or decreasing. Furthermore, by parameterising the covariance rather than systematics signal directly, the GP model is capable of marginalising over a broader range of function space with relatively few tunable parameters.

Before proceeding, we highlight the fact that the covariance kernels given by Equations \ref{eq:ch36_ch45_kernel}--\ref{eq:ch58_ch80_kernel} assume that the input variables are noise-free. While this is a reasonable assumption for time $t$, it is not necessarily the case for the centroid coordinates $x$ and $y$. Indeed, the undersampled nature of the IRAC PSF makes the centroid estimates particularly susceptible to shot noise of individual pixels. However, due to the fact that this noise is white, and because in practise a small patch of a single pixel is densely sampled by the PSF over the course of a few hours, we expect its effect to be averaged out. For this reason, and in line with other analyses of IRAC lightcurves, we do not explicitly account for the noise of $x$ and $y$. In future work, however, it may be worth considering a more explicit treatment of noisy inputs \citep[e.g.][]{Goldberg98regressionwith, NIPS2011_4295}, or simply smoothing noisy inputs before feeding them to the GP \citep[e.g.][]{2012MNRAS.419.2683G}.

\subsection{Lightcurve binning} \label{sec:lcbinning}

GP models become computationally intractable for datasets with $N \gtrsim 10^3$ data points. This is due to the need to factorise the $N \times N$ covariance matrix when evaluating the $-\frac{1}{2}\mb{r}^{\tn{T}}(\mb{K}+\gb{\Sigma})^{-1}\mb{r}$ term and computing the determinant $-\frac{1}{2}\ln|\mb{K}+\gb{\Sigma}|$ for each log likelihood evaluation (Equation \ref{eq:logp_gp_general_expanded}). Our code implements this using Cholesky factorisation, which has a computational cost scaling as $\bigO\left(\,N^3\,\right)$. To apply GPs to IRAC datasets, most of which consist of $N > 10^4$ data points, we bin the fluxes and centroid $xy$ coordinates in time prior to fitting. Binning factors were chosen according to the format of individual lightcurves, such that the time interval between successive binned points was $\lesssim 15$\,sec and the number of binned points per lightcurve was $N_b=1000\text{--}1500$. Lightcurves obtained in full array mode for Program 461 were binned by a factor of two, giving a median cadence of about 17\,sec; lightcurves obtained in subarray mode for Programs 40280 and 60021 with frame times of 0.4\,sec were binned by a factor of 32, giving a median cadence of about 14\,sec; lightcurves obtained in subarray mode for Program 60021 with frame times of 0.1\,sec were binned by a factor of 128, giving a median cadence of about 17\,sec; lightcurves obtained in subarray mode for Program 20523 with frame times of 0.1\,sec were only binned by a factor of 32, due to the sparser sampling of the lightcurve as the telescope was constantly repointed during the observations, resulting in a median cadence of about 5\,sec. Details are given in Table \ref{table:irac_lightcurves}.

The obvious drawback of binning the lightcurves in time is that we lose information on the timescales of our bin sizes. For instance, the $xy$ centroid coordinates can vary coherently over timescales $\lesssim 15$\,sec \citep[e.g.][]{2012ApJ...754..136S}. However, this should not affect our results significantly, as none of the astrophysical quantities of interest vary over the bin timescales, nor does the information content of the transit lightcurve degrade significantly as we reduce the time resolution to $\lesssim 15$\,sec. Furthermore, high-frequency systematics should mostly average out given the large number of binned data points for each tunable model parameter. Any correlations that remain will be accounted for by the time-dependent Mat\'{e}rn $\nu = 3/2$ kernel $k_t$ in our model (Equations \ref{eq:ch36_ch45_kernel} and \ref{eq:ch58_ch80_kernel}).

\section{Lightcurve fitting} \label{sec:lcfitting}

Lightcurves were fit individually by marginalising over the model posterior distribution using Markov chain Monte Carlo (MCMC), in order to quantify the degeneracies between the astrophysical parameters of interest and instrumental systematics. Following Bayes theorem, the model posterior distribution is given by $p(\gb{\alpha},\gb{\gamma}|\mb{d})  \propto p(\mb{d}|\gb{\alpha},\gb{\gamma})\,p(\gb{\alpha})\,p(\gb{\gamma})$, where $p(\mb{d}|\gb{\alpha},\gb{\gamma})$ is the GP likelihood given by Equation \ref{eq:logp_gp_general_expanded}, and $p(\gb{\alpha})$ and $p(\gb{\gamma})$ are the priors on the mean function and covariance parameters, respectively. We adopted uniform priors for the mean function parameters $\gb{\alpha}$ and covariance length scales $L_i$. For the covariance amplitudes $A_i$, Gamma distribution priors of the form $p(A_i)=\tn{Gam}(1,100)\propto \exp[ -100 A_i ]$ for $i=\{t,xy,\tau\}$ were adopted. The latter give decreasing probability to increasing covariance amplitudes, encouraging the GP to reduce the covariance amplitude unless justified by the data. The white noise level $\sigma_w$ was included as a free parameter in each model, with a uniform prior. The ability to inflate the statistical errorbars above the formal shot noise floor provides the models with some additional flexibility for dealing with high-frequency noise that may be present in the data, without having to reduce the correlation length scales $L_i$ of the covariance kernels to unreasonably small values. 

With the posterior distributions defined, the model fitting for each lightcurve proceeds as follows. Values for the model parameters were drawn randomly from the model prior, i.e.\ $p(\alphab)\,p(\gammab)$. With this as a starting point, Equation \ref{eq:logp_gp_general_expanded} was optimised using the Nelder-Mead simplex algorithm \citep{neldermead_1965} to obtain maximum likelihood estimates (MLEs) for the parameters. A short Metropolis-Hastings MCMC chain \citep{MetropolisRosenbluth53, hastings70} of 1000 steps was initiated at the MLE, with step sizes pretuned to give acceptance rates of 20--40\,\%. The median chain values were then used as the starting location for a second MLE optimisation. In practise, the randomness introduced by the short MCMC chains helped prevent the MLE optimisations from getting trapped at local maxima of the likelihood surface, thus increasing the chance of locating the global likelihood maximum. To further increase this probability, the entire process was repeated ten times, each time from a different random starting point. This was done separately for each lightcurve produced using the different photometric aperture sizes (Section \ref{sec:datared}). The photometric reduction giving the lowest scatter in the residuals was then selected for the remaining analysis.

Before commencing the final MCMC chains, the covariance parameters were fixed to their MLE values, which are reported in Table \ref{table:covpar_mle}. This approach -- which is often referred to as ``type-II maximum likelihood'' \citep[for further discussion see][]{2012MNRAS.419.2683G} -- allows the expensive $\bigO\left(\,N^3\,\right)$ covariance matrix factorisation required for the GP likelihood evaluation (Equation \ref{eq:logp_gp_general_expanded}) to be performed only once at the beginning of the chain. Subsequent steps only cost $\bigO\left(\,N^2\,\right)$, resulting in much faster computations. The disadvantage is that by fixing the covariance parameters $\gb{\gamma}$, they are not marginalised over. In effect, this imposes an artificial restriction on the range of systematics functions that are explored by the GP model. Consequently, there may be degeneracies between the planet signal and systematics that are not fully incorporated into the final uncertainties for the planet parameters $\gb{\alpha}$ presented here. For example, in their re-analysis of the NICMOS transmission spectrum for HD\,189733b, \cite{2012MNRAS.419.2683G} found uncertainties that were up to $\sim 1.5$ larger when the covariance parameters were allowed to vary in the marginalisation compared to when they were fixed to their MLE values. Therefore, the uncertainties we report in this study should be considered lower limits to the true uncertainties. 

It should be emphasised, however, that fixing the covariance parameters is quite different to fixing the parameters of an explicit functional model for the systematics. Instead, fixing the covariance parameters is somewhat analogous to selecting a family of parametric models for the systematics, as they control the high-level properties of the function space spanned by the GP model. Rather than selecting from a handful of distinct parametric models, the GP model offers access to a continuum of possible functions. By using covariance parameters that optimise the GP likelihood, this continuum is narrowed in a principled manner. To compare with the bilinear interpolation method for pixel mapping used by \cite{2012ApJ...754..136S} for instance, optimising the covariance length scales $L_x$ and $L_y$ is similar to choosing the optimal grid spacing for the interpolation knots. Treating our dataset as a GP, we have the advantage of being able to do this in the context of a self-consistent probabilistic model, by maximising the likelihood function with respect to the unknown parameters using a numerical optimiser.

Having fixed the covariance parameters, an initial chain of $10^5$ steps was run with the planet parameters allowed to vary and step sizes again pretuned to ensure acceptance rates of 20--40\,\%. The first $5 \times 10^4$ steps were discarded as burn-in. An additional four chains were then run for $10^5$ steps each, with starting parameter values drawn randomly from normal distributions centred on the mean values of the first chain. The width of the normal distributions were taken to be five times the standard deviation of the first chain, to ensure the starting locations were well-dispersed in parameter space. After discarding the first $5 \times 10^4$ burn-in steps of these chains, the Gelman-Rubin statistics for each parameter were calculated \citep{GelmanRubin92}. For all lightcurves, these were found to be well within 1\,\% of unity, consistent with the chains having reached stable states. Finally, the five independent chains were combined into a single chain, giving $2.5 \times 10^5$ samples from the posterior distribution.

\section{Results} \label{sec:results}

\begin{table*}
\begin{minipage}{\linewidth}
\centering
\caption{Results of MCMC primary transit lightcurve analyses. Quoted values are the chain medians, with uncertainties corresponding to the ranges either side of the medians that contain 34\% of the chain samples. Orbital inclination values $i$ are derived from the corresponding impact parameter $b=a\cos i/\Rs$ and normalised semimajor axis $\aRs$ values. Brightness temperatures $T_b$ are derived from the measured eclipse depths $\FpFs$ assuming an ATLAS stellar model \citep{1979ApJS...40....1K, 1993KurCD..13.....K} for HD\,209458 and integrating over the IRAC bandpasses. \label{table:mcmc_results}}
\begin{tabular}{ccccccc} 
\scriptsize \\
\hline \\
\multicolumn{7}{c}{Transits}  \medskip \\
\hline \\
Channel & Date & $T_{\textnormal{mid}}$ & $R_p/R_\star$ & $a/R_\star$ & $b$ & $i$ \\
($\mu$m) & & ($\textnormal{BJD}_{\textnormal{UTC}}-2450000$) & & & & (deg) \smallskip \\
\hline \\

3.6 & 2007 Dec 31 & ${4465.63711}_{-0.00027}^{+0.00027}$ & ${0.12077}_{-0.00084}^{+0.00085}$ & ${8.72}_{-0.23}^{+0.26}$ & ${0.527}_{-0.044}^{+0.035}$ & ${86.54}_{-0.33}^{+0.39}$ \smallskip \\ 
    & 2008 Jul 19 & ${4666.54742}_{-0.00021}^{+0.00021}$ & ${0.12220}_{-0.00062}^{+0.00062}$ & ${8.89}_{-0.26}^{+0.33}$ & ${0.482}_{-0.062}^{+0.044}$ & ${86.89}_{-0.39}^{+0.50}$ \smallskip \\ 
    & 2011 Jan 14 & ${5575.93102}_{-0.00030}^{+0.00030}$ & ${0.11354}_{-0.00087}^{+0.00085}$ & ${8.77}_{-0.33}^{+0.33}$ & ${0.525}_{-0.056}^{+0.048}$ & ${86.57}_{-0.46}^{+0.48}$ \smallskip \\ 
    & 2014 Feb 15 & ${6703.85250}_{-0.00011}^{+0.00011}$ & ${0.11919}_{-0.00032}^{+0.00032}$ & ${8.13}_{-0.10}^{+0.10}$ & ${0.590}_{-0.014}^{+0.013}$ & ${85.84}_{-0.14}^{+0.14}$ \medskip \\ 
4.5 & 2008 Jul 22 & ${4670.07290}_{-0.00029}^{+0.00029}$ & ${0.12199}_{-0.00091}^{+0.00094}$ & ${9.31}_{-0.30}^{+0.35}$ & ${0.423}_{-0.079}^{+0.057}$ & ${87.39}_{-0.45}^{+0.56}$ \smallskip \\ 
    & 2010 Jan 19 & ${5216.40564}_{-0.00007}^{+0.00007}$ & ${0.12099}_{-0.00029}^{+0.00029}$ & ${8.89}_{-0.07}^{+0.06}$ & ${0.493}_{-0.010}^{+0.011}$ & ${86.82}_{-0.10}^{+0.08}$ \medskip \\ 
5.8 & 2007 Dec 31 & ${4465.63663}_{-0.00031}^{+0.00032}$ & ${0.12007}_{-0.00265}^{+0.00248}$ & ${9.20}_{-0.35}^{+0.36}$ & ${0.435}_{-0.082}^{+0.066}$ & ${87.29}_{-0.53}^{+0.59}$ \smallskip \\ 
   & 2008 Jul 19 & ${4666.54744}_{-0.00033}^{+0.00033}$ & ${0.11880}_{-0.00272}^{+0.00284}$ & ${9.27}_{-0.37}^{+0.38}$ & ${0.427}_{-0.091}^{+0.071}$ & ${87.36}_{-0.56}^{+0.64}$ \medskip \\ 
8.0 & 2007 Dec 24 & ${4458.58730}_{-0.00013}^{+0.00013}$ & ${0.12007}_{-0.00114}^{+0.00114}$ & ${8.78}_{-0.11}^{+0.11}$ & ${0.513}_{-0.019}^{+0.017}$ & ${86.65}_{-0.15}^{+0.16}$ \smallskip \\ 
   & 2008 Jul 22 & ${4670.07243}_{-0.00022}^{+0.00022}$ & ${0.11991}_{-0.00073}^{+0.00073}$ & ${8.67}_{-0.18}^{+0.20}$ & ${0.527}_{-0.032}^{+0.027}$ & ${86.52}_{-0.26}^{+0.28}$ \medskip \\ %\\ \hline 
\hline \\
\multicolumn{5}{c}{Eclipses}  \medskip \\
\hline \\

Channel & Date & $T_{\textnormal{mid}}$ & $\FpFs$ & $T_b$  \\
($\mu$m) & & ($\textnormal{BJD}_{\textnormal{UTC}}-2450000$) & (\%) & (K)  \smallskip \\
\hline \\
3.6 & 2005 Nov 28 & ${3702.52741}_{-0.00457}^{+0.00476}$ & ${0.093}_{-0.034}^{+0.033}$ & ${1447}_{-165}^{+163}$ \smallskip \\ 
    & 2011 Jan 12 & ${5574.16955}_{-0.00190}^{+0.00173}$ & ${0.122}_{-0.011}^{+0.011}$ & ${1591}_{-49}^{+49}$ \smallskip \\ 
    & 2011 Jan 16 & ${5577.69695}_{-0.00193}^{+0.00204}$ & ${0.124}_{-0.014}^{+0.014}$ & ${1599}_{-65}^{+65}$ \smallskip \\ 
    & 2014 Feb 13 & ${6702.09332}_{-0.00400}^{+0.00397}$ & ${0.112}_{-0.017}^{+0.016}$ & ${1545}_{-80}^{+76}$ \smallskip \\ 
    & 2014 Feb 17 & ${6705.61556}_{-0.00128}^{+0.00136}$ & ${0.106}_{-0.008}^{+0.007}$ & ${1515}_{-37}^{+34}$ \medskip \\ 
4.5 & 2005 Nov 28 & ${3702.52942}_{-0.00191}^{+0.00169}$ & ${0.145}_{-0.017}^{+0.018}$ & ${1474}_{-69}^{+72}$ \smallskip \\ 
    & 2010 Jan 18 & ${5214.64693}_{-0.00079}^{+0.00085}$ & ${0.140}_{-0.014}^{+0.014}$ & ${1455}_{-56}^{+57}$ \smallskip \\ 
    & 2010 Jan 21 & ${5218.16915}_{-0.00138}^{+0.00147}$ & ${0.133}_{-0.011}^{+0.011}$ & ${1426}_{-48}^{+47}$ \medskip \\ 
5.8 & 2005 Nov 28 & ${3702.53033}_{-0.00664}^{+0.00457}$ & ${0.142}_{-0.058}^{+0.059}$ & ${1297}_{-220}^{+225}$ \medskip \\ 
8.0 & 2005 Nov 28 & ${3702.53461}_{-0.00344}^{+0.00322}$ & ${0.225}_{-0.063}^{+0.064}$ & ${1433}_{-216}^{+219}$ \smallskip \\ 
   & 2007 Dec 25 & ${4460.35170}_{-0.00107}^{+0.00100}$ & ${0.215}_{-0.012}^{+0.012}$ & ${1397}_{-43}^{+41}$ \\ \\ \hline 
\end{tabular}
\end{minipage}
\end{table*}

\begin{figure*}
\centering  % this centres figure in column
\includegraphics[width=\linewidth]{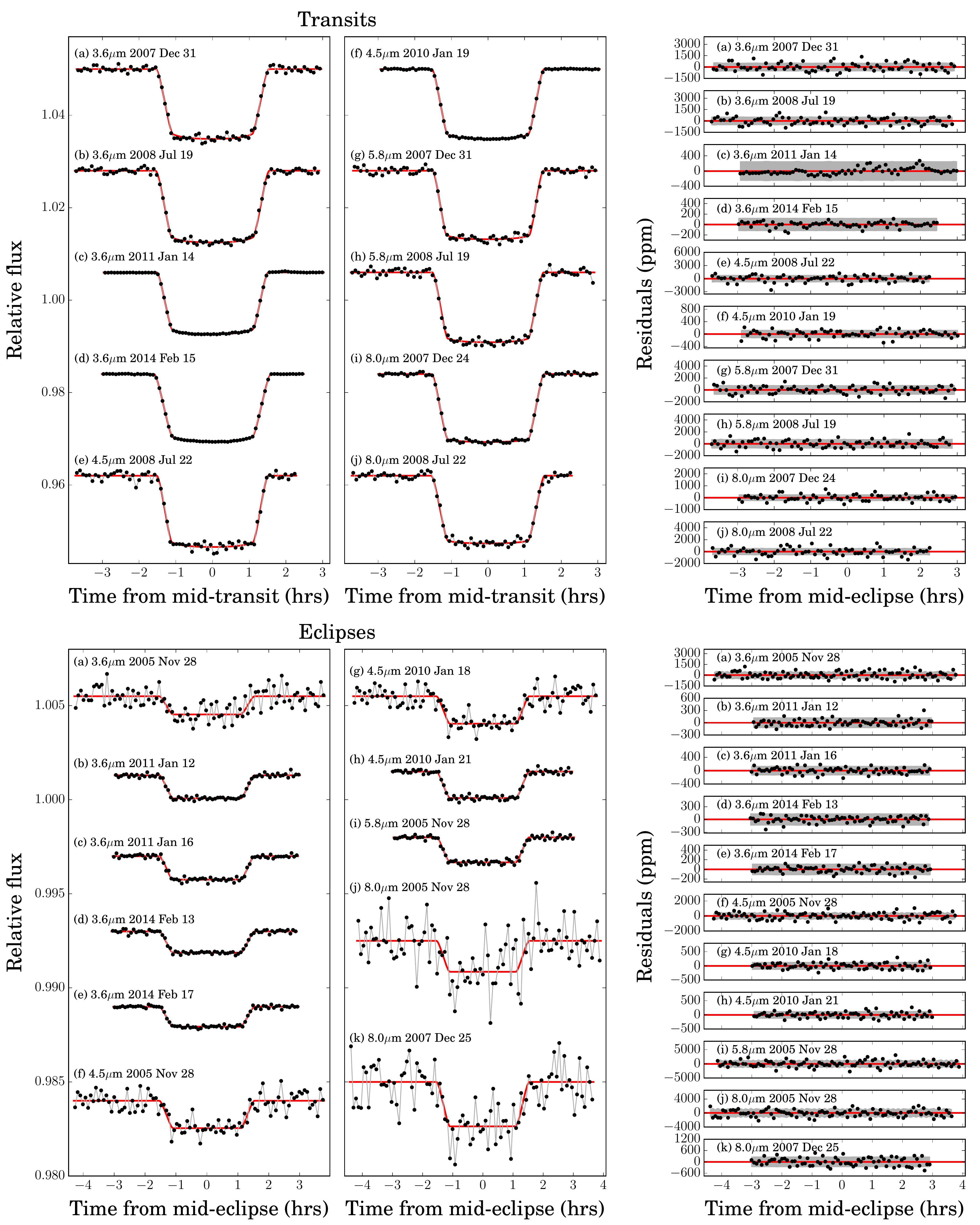}
\caption{Corrected primary transit (top) and secondary eclipse (bottom) lightcurves obtained by dividing the raw lightcurves by the systematics components of the best-fit GP models. Model residuals are shown in the rightmost column, with grey shaded regions indicating the inferred white noise values $\sigma_w$.}
\label{fig:lcs_corr}
\end{figure*}

Results of the primary transit and secondary eclipse MCMC analyses are given in Table \ref{table:mcmc_results}. Best-fit models are overplotted on the raw lightcurves in Figure \ref{fig:lcs_raw} and the corrected lightcurves with residuals are shown in Figure \ref{fig:lcs_corr}. 

\subsection{Transmission ($\RpRs$)} \label{sec:results_rprs}

We find good agreement in the inferred parameters across epochs for the majority of the transit lightcurves. The only exceptions to this are the 2011 Jan 14 and 2014 Feb 15 transits measured in the $3.6\um$ channel. For the 2011 lightcurve we obtain a value for $\RpRs$ that is $>6\sigma$ discrepant relative to those obtained for the 2007 Dec 31 and 2008 Jul 19 $3.6\um$ lightcurves. This lightcurve has been classified as a failed observation by the \textit{Spitzer} Science Center due to the presence of high-frequency noise of unknown origin during the second half of the transit. For the 2014 lightcurve, we obtain values for $\aRs$ and $b$ that are both $\sim 2\sigma$ discrepant relative to the values inferred for the 2007 and 2008 lightcurves, while the value inferred for $\RpRs$ is $>4\sigma$ discrepant. The source of this disagreement is not clear. We also performed the lightcurve fitting using polynomial $xy$ decorrelations and the PLD method of \cite{2014arXiv1411.7404D}, but these analyses gave similarly discrepant results. We therefore suspect that there is either an issue with the data itself or our photometric reduction for this particular lightcurve. For these reasons, we do not consider the 2011 and 2014 $3.6\um$ transit lightcurves any further in this paper. However, we consider our analyses for the eclipses in these lightcurves to be more robust, as they gave results that are consistent with those obtained at other epochs.

\begin{figure*}
\centering  % this centres figure in column
\includegraphics[width=\linewidth]{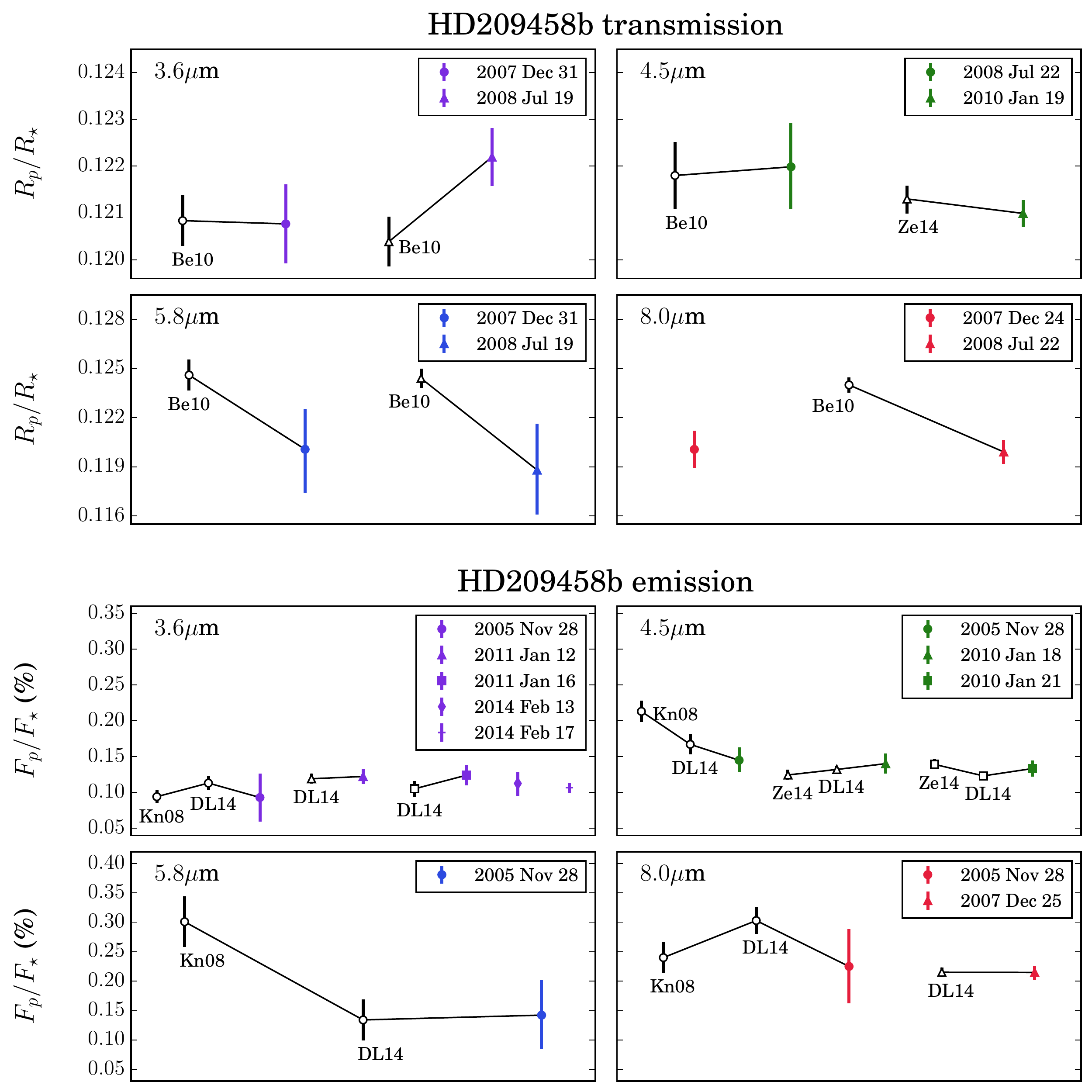}
\caption{Comparison of transmission $\RpRs$ (top) and emission $\FpFs$ (bottom) results obtained in the present study with those published in the literature. Wavelength channels are indicated in the top left corner of each axis. Independent analyses of the same dataset are linked by solid black lines:\ filled symbols show the values obtained in the current study and unfilled symbols show values obtained by other authors. The latter are labelled using the same abbreviations adopted in Table \ref{table:irac_datasets}.}
\label{fig:RpRs_and_FpFs_epoch_comparisons}
\end{figure*}

For the remaining eight transit lightcurves, inferred values for $\RpRs$ are shown in Figure \ref{fig:RpRs_and_FpFs_epoch_comparisons}, along with values previously published in the literature. In the $3.6\um$ channel, we find  $\RpRs = {0.12077}_{-0.00084}^{+0.00085}$ for the 2007 Dec 31 lightcurve, which is in agreement with the value of $\RpRs = 0.120835 \pm 0.00054$ reported by \cite{2010MNRAS.409..963B}. We obtain a somewhat higher value of $\RpRs = {0.12220}_{-0.00062}^{+0.00062}$ for the 2008 Jul 19 lightcurve, compared with the value of $\RpRs = 0.120387 \pm 0.00053$ obtained by Beaulieu et al. 

Our results for the $4.5\,\um$ lightcurves are in good agreement both with each other and with the values previously published by \cite{2010MNRAS.409..963B} and \cite{2014ApJ...790...53Z}. 

For the $5.8\,\um$ 2007 Dec 31 and 2008 Jul 19 lightcurves, we find $\RpRs=0.12007_{-0.00265}^{+0.00248}$ and $\RpRs=0.11880_{-0.00272}^{+0.00284}$, respectively. Although these values are consistent with each other, they are $1.7\sigma$ and $1.9\sigma$ lower, respectively, than the corresponding values of $\RpRs = 0.1246 \pm 0.00095$ and $\RpRs = 0.1244 \pm 0.00059$ obtained by \cite{2010MNRAS.409..963B}. In particular, our uncertainties are $\sim 2.5$--$3$ times larger than those of Beaulieu et al. This is most likely due to the flexibility of the GP models allowing broader ranges of function space to be marginalised over, which in turn maps out broader degeneracies between the transit parameters and the systematics contributions.

For the $8.0\um$ channel, we find good agreement between our inferred parameters for the 2007 Dec 24 and 2008 Jul 22 lightcurves, with $\RpRs = {0.12007}_{-0.00114}^{+0.00114}$ and $\RpRs = {0.11991}_{-0.00073}^{+0.00073}$, respectively. For the 2008 lightcurve, our value is $4.7\sigma$ lower than the value of $\RpRs = 0.1240 \pm 0.00046$ published by \cite{2010MNRAS.409..963B}.

\subsection{Emission ($\FpFs$)}

As with the transmission measurements, we find consistent results across epochs and wavelength channels for the inferred emission values (Figure \ref{fig:RpRs_and_FpFs_epoch_comparisons}). Our results are mostly in agreement with values previously published in the literature, but with typically larger uncertainties. There are two notable exceptions. Firstly, for the $4.5\um$ 2005 Nov 28 lightcurve, we obtain $\FpFs = {0.145}_{-0.017}^{+0.018}$\%, which is $2.9\sigma$ lower than the value of $\FpFs = 0.213 \pm 0.015$\% published by \cite{2008ApJ...673..526K}, but in agreement with the value of $\FpFs = 0.134 \pm 0.035$\% published by \cite{2014ApJ...796...66D} for the same lightcurve. Our revision brings the value into good agreement with those obtained in the same wavelength channel at different epochs, both in the current study and by \cite{2014ApJ...790...53Z}. Secondly, for the 2005 Nov 28 $5.8\um$ lightcurve, we obtain $\FpFs = {0.142}_{-0.058}^{+0.059}$\%, which is $2.3\sigma$ lower than the value of $\FpFs = 0.310 \pm 0.043$\% published by \cite{2008ApJ...673..526K}, but in agreement with the value of $\FpFs = 0.134 \pm 0.035$\% published by \cite{2014ApJ...796...66D}. Finally, we note that unlike \cite{2014ApJ...796...66D}, who obtained $\FpFs$ values for the 2005 Nov 28 and 2007 Dec 25 $8.0\um$ lightcurves that conflicted at the $3.6\sigma$ level, our values are in good agreement with each other due to the relatively large uncertainty we obtain for the 2005 Nov 28 eclipse depth. We therefore favour the more conservative uncertainty estimate provided by our GP analysis in this case.

\subsection{Orbital parameters ($\aRs$, $b$)}

\begin{figure}
\centering  % this centres figure in column
\includegraphics[width=\linewidth]{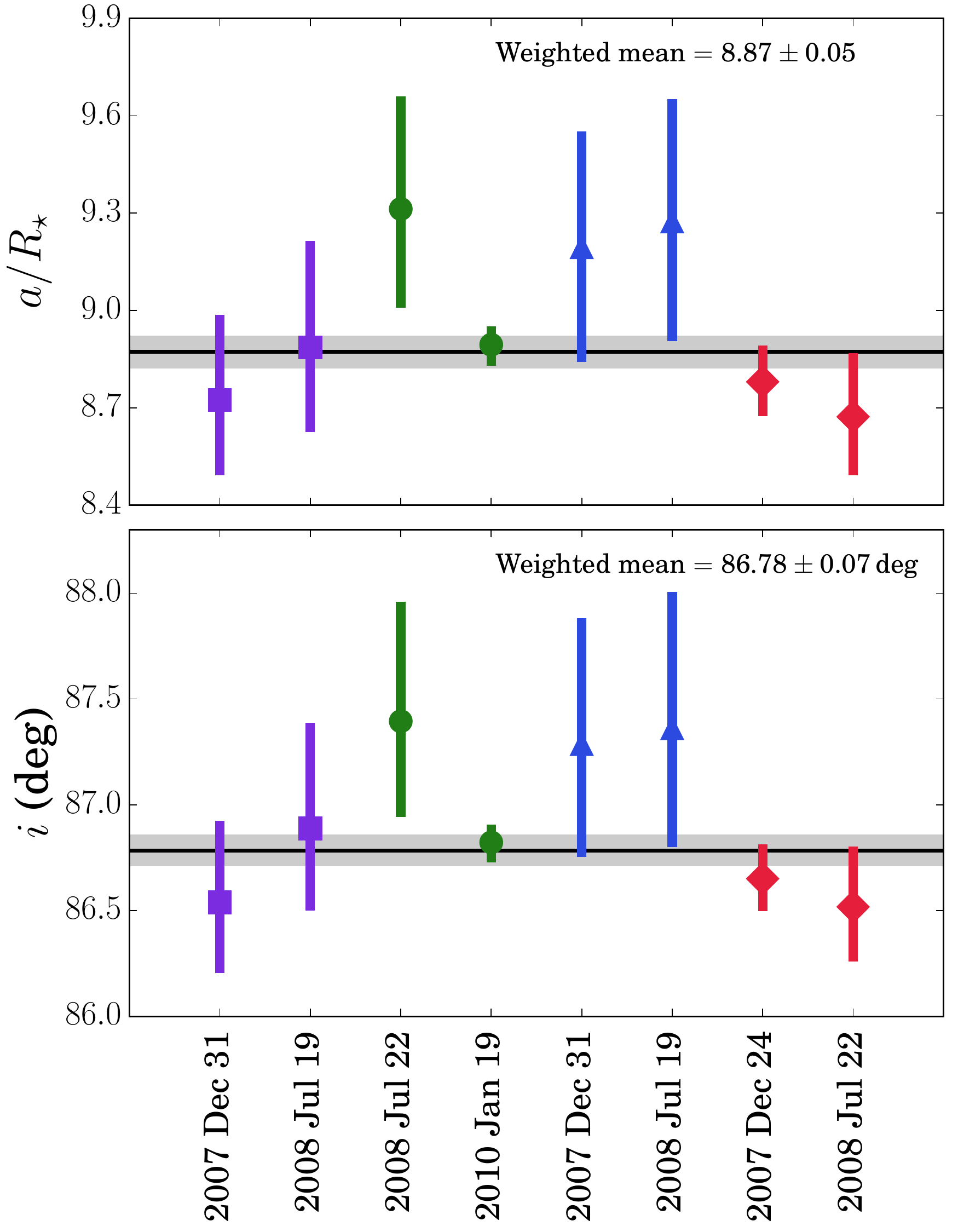}
\caption{Normalised semimajor axis $\aRs$ and orbital inclination values obtained from the MCMC analyses. Lightcurve epochs are labelled along the horizontal axis, with different marker symbols for each wavelength channel:\ purple squares for $3.6\um$, green circles for $4.5\um$, blue triangles for $5.8\um$ and red diamonds for $8.0\um$. Weighted mean values are labelled on each axis, and shown as horizontal black lines with grey shaded regions indicating the corresponding $1\sigma$ uncertainties.}
\label{fig:aRs_and_incl_epoch_comparisons}
\end{figure}

The normalised semimajor axis $\aRs$ and orbital inclination $i$ values recovered from the primary lightcurve analyses are plotted in Figure \ref{fig:aRs_and_incl_epoch_comparisons}. Note the clear correlation between both parameters, i.e.\ higher values of $\aRs$ are associated with higher values of $i$, and vice versa. This reflects the fact that these parameters exert opposing influences on the transit duration.

We expect $\aRs$ and $i$ remain constant in time across the different wavelength channels. Computing the weighted arithmetic mean across epochs, we obtain $\aRs = 8.87 \pm 0.05$, $b = 0.499 \pm 0.008$ and $i = 86.78 \pm 0.07 $\,deg. We caution that the quoted uncertainties for the weighted means are likely to be underestimated, as they are calculated by combining multiple measurements under the assumption that the errorbars are normally distributed, which is not necessarily true. Nonetheless, these values are consistent at the $\sim 1\sigma$ level with the values of $\aRs = 8.77 \pm 0.07$ and $i=86.76 \pm 0.10$\,deg obtained by \cite{2010MNRAS.409..963B}, and $\aRs = 8.810_{-0.069}^{+0.064}$ and $i = 86.69_{-0.10}^{+0.09}$\,deg obtained by \cite{2014ApJ...790...53Z}. 

Indeed, the values obtained for $\aRs$ and $b = a/\Rs\cos i$ in the present study are overall more consistent across wavelengths and epochs than previously published values. For instance, the $\aRs$ and $i$ values reported by \cite{2010MNRAS.409..963B} for the $8.0\,\um$ channel differ from most of the values they obtain in the other channels by $3\mbox{--}5\sigma$. The relative consistency of the orbit parameters derived in the current study therefore offers further evidence that the GP modelling approach is doing a good job of accounting for the lightcurve systematics and providing realistic parameter uncertainties.

\subsection{Ephemeris} \label{sec:ephemeris}

\begin{table*}
\begin{minipage}{\linewidth} 
\centering
\caption{Results of the ephemeris fit to transit and eclipse mid-times assuming a constant orbital period and zero eccentricity. As discussed in Section \ref{sec:ephemeris}, fits were performed both with and without the 2010 Jan 19 transit mid-time. As discussed in the text, we favour the ephemeris obtained with the 2010 Jan 19 transit excluded from the fit. \label{table:ephemeris}}
\begin{tabular}{cccccc} 
\hline \\
 &  &  & \multicolumn{3}{c}{$O-C$ residual (min)} \medskip \\ \cline{4-6}
\\
Date & Signal & Channel ($\um$) & Including 2010 transit && Excluding 2010 transit \smallskip \\ 
\hline \\

2005 Nov 28 & Eclipse & 3.6 & $-6.6 \pm 6.7$  &&  $-1.8 \pm 6.7$ \\
            &         & 4.5 & $-3.7 \pm 2.6$  &&  $1.1 \pm 2.7$ \\
            &         & 5.8 & $-2.4 \pm 8.1$  &&  $2.4 \pm 8.1$ \\
            &         & 8.0 & $3.8 \pm 4.8$ && $8.6 \pm 4.8$ \smallskip \\
2007 Dec 24 & Eclipse & 8.0 & $-0.5 \pm 0.2$ &&  $-0.1 \pm 0.2$ \\
2007 Dec 25 & Transit & 8.0 & $2.4 \pm 1.5$ && $2.8 \pm 1.5$ \smallskip \\
2007 Dec 31 & Transit & 3.6 & $-0.1 \pm 0.4$ && $0.3 \pm 0.4$ \\
            &         & 5.8 & $-0.7 \pm 0.5$  &&  $-0.4 \pm 0.5$ \smallskip \\
2008 Jul 19 & Transit & 3.6 & $0.5 \pm 0.3$ && $-0.3 \pm 0.3$  \\
            &         & 5.8 & $0.6 \pm 0.5$ && $-0.2 \pm 0.5$  \smallskip \\
2008 Jul 22 & Transit & 4.5 & $1.6 \pm 0.4$ && $0.8 \pm 0.4$  \\
            &         & 8.0 & $0.9 \pm 0.3$ && $0.1 \pm 0.3$  \smallskip \\
2010 Jan 18 & Eclipse & 4.5 & $5.1 \pm 1.2$ && $1.1 \pm 1.3$  \\
2010 Jan 19 & Transit & 4.5 & $-0.2 \pm 0.2$ && $-4.7 \pm 0.5$ \\
2010 Jan 21 & Eclipse & 4.5 & $1.5 \pm 2.1$ &&  $-2.5 \pm 2.1$ \\
2011 Jan 12 & Eclipse & 3.6 & $4.6 \pm 2.6$  &&  $-1.5 \pm 2.7$ \\
2011 Jan 16 & Eclipse & 3.6 & $8.4 \pm 2.9$ &&  $2.3 \pm 3.0$ \\
2014 Feb 13 & Eclipse & 3.6 & $16.2 \pm 5.8$ &&  $3.7 \pm 5.9$ \\
2014 Feb 17 & Eclipse & 3.6 & $12.6 \pm 2.0$ &&  $0.1 \pm 2.3$ \smallskip \\ \hline 
\rule{0pt}{4ex} & \multicolumn{2}{r}{Period (day)} & $ 3.5247361 \pm 6 \times 10^{-7}$ && $3.524750 \pm 1 \times 10^{-6}$\smallskip \\
\multicolumn{3}{r}{$T_0$ (BJD$_{\tn{TDB}}$)} & $2454560.80567 \pm 8 \times 10^{-5}$ && $2454560.80588 \pm 8 \times 10^{-5}$ \medskip \\  \hline

\end{tabular}
\end{minipage}
\end{table*}

Using the transit and eclipse mid-times listed in Table \ref{table:mcmc_results}, we computed the ephemeris assuming a constant orbital period $P$, with the linear relation $T_0 = T_{\tn{mid}}(n) + n P$, where $n$ is the number of orbital periods since the reference epoch $T_0$. Eclipse mid-times were treated as occurring precisely $0.5P$ after the immediately-preceding transit. Before performing the fit, we converted the BJD timestamps from Coordinated Universal Time (UTC) to the Barycentric Dynamical Time (TDB) standard, as recommended by \cite{2010PASP..122..935E}. To do this, we added the appropriate number of leap seconds to the UTC timestamps, namely:\ 64.184\,sec for the 2005 lightcurves; 65.184\,sec for the 2007--2008 lightcurves; 66.184\,sec for the 2010--2011 lightcurves; and 67.184\,sec for the 2014 lightcurves. We report the results in Table \ref{table:ephemeris}. 

When all of the mid-times are included in the fit, we obtain $T_0 = 2454560.80567 \pm 8 \times 10^{-5}$ BJD$_{\tn{TDB}}$ and $P = 3.5247361 \pm 6 \times 10^{-7}$\,day, with a reduced $\chi^2=7.0$ indicating a poor overall fit to the data. However, we found that when we exclude the 2010 Jan 19 transit mid-time, we obtain $T_0 = 2454560.80588 \pm 8 \times 10^{-5}$ BJD$_{\tn{TDB}}$ and $P = 3.524750 \pm 1 \times 10^{-6}$\,day, with a much improved reduced $\chi^2=1.0$. It is not clear why our 2010 Jan 19 transit fails to fit a linear ephemeris, although we note that our measured mid-time for this lightcurve is in excellent agreement at the $0.01\sigma$ level with the value obtained by \cite{2014ApJ...790...53Z}. A full investigation into the cause of this discrepancy is beyond the scope of the current paper. However, given that the other eighteen mid-times are well fit by a linear ephemeris, we conclude that the data is consistent with a constant orbital period.

\section{Discussion} \label{sec:discussion}

The results outlined in Section \ref{sec:results} demonstrate the effectiveness of the GP modelling approach for handling systematics in IRAC lightcurves. Compared with those previously published in the literature, the planet properties inferred from the GP analyses are overall more consistent across different epochs and, for the wavelength-independent properties, across the different wavelength channels. In many cases, this is due to the GP analyses giving uncertainty estimates that are up to $\sim 4$ times larger than those reported by other authors. For the reasons given in Section \ref{sec:gpmodels}, we argue that the GP uncertainties provide a more realistic reflection of our ignorance. This is primarily because the GP models offer greater flexibility for handling systematics that do not have a well-understood functional form, compared with the simple parametric approximations used widely in the literature. Marginalisation of the GP model posterior distributions therefore allows us to more exhaustively explore possible degeneracies between the planet signal and systematics, and incorporate these into the uncertainties associated with the inferred planet properties. However, recall from Section \ref{sec:lcfitting} that we fixed the covariance parameters to their MLE values before marginalising over the planet parameters with MCMC, which means the uncertainties quoted in Table \ref{table:mcmc_results} should in fact be regarded as lower limits to the true uncertainties. 

Further verification of the reliability of GP models for inferring planet parameters from IRAC lightcurves could be obtained by systematically applying the method to synthetic datasets, similar to what was done by \cite{2014MNRAS.445.3401G}. This would allow us to directly compare inferred values for the planet parameters with their known values. The challenge with such an approach, of course, is that it requires a realistic simulation of the IRAC systematics, for which we do not possess a functional form. With this caveat in mind, however, such an investigation would be useful in the future.

\subsection{Atmosphere implications} \label{sec:atmosphere_implications}

\begin{figure*}
\centering  % this centres figure in column
\includegraphics[width=\linewidth]{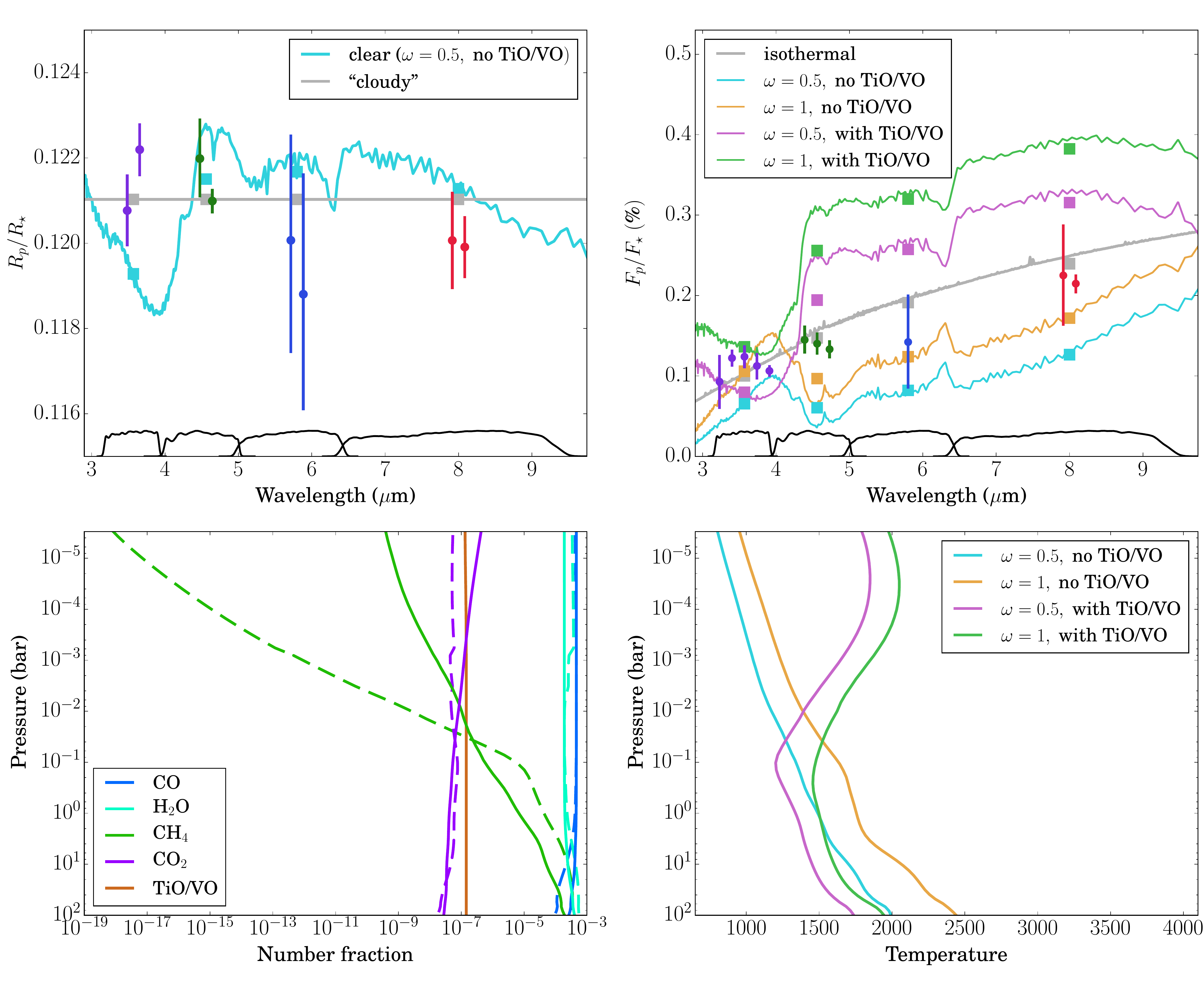}
\caption{Top panels show the transmission $\RpRs$ (left) and emission $\FpFs$ (right) measurements obtained in the current study for HD\,209458b, with colours indicating wavelength channels. IRAC bandpasses are indicated at the bottom of both axes as solid black lines. For measurements obtained in the same wavelength channel, small horizontal offsets have been applied for clarity. Both panels show 1D \texttt{ATMO} models assuming solar abundances and chemical equilibrium. Corresponding chemical abundances are shown in the lower left panel and pressure-temperature profiles are shown in the lower right panel. For the abundances, solid lines correspond to uniform heat redistribution from the dayside to the nightside (i.e.\ $\omega=0.5$), while dashed lines correspond to zero heat redistribution (i.e.\ $\omega=1$). For the transmission, the light blue line shows the clear atmosphere obtained assuming $\omega=0.5$, while the grey line shows a simple model corresponding to an opaque cloud deck at $\RpRs=0.1210$, based on fitting a simple cloud model to the data as described in Section \ref{sec:atmosphere_implications}. For the emission, clear atmosphere models are shown for $\omega=0.5$ and $\omega=1$, both with and without TiO/VO in the atmosphere. For those models including TiO/VO, the species were assumed to be mixed uniformly throughout the atmosphere, as shown by the brown line in the lower left panel. The inclusion of TiO/VO results in inverted temperature-pressure profiles, as shown in the lower right panel. Also shown in the upper right panel is the emission spectrum obtained using an ATLAS stellar model for HD\,209458 and assuming the planet radiates as a blackbody with a temperature of 1484\,K. In the top panels, square symbols give the model values integrated over the corresponding IRAC bandpasses.}
\label{fig:transmission_and_emission_spectra}
\end{figure*}

The IRAC data analysed in the current study address two fundamental hypotheses concerning the nature of HD\,209458b's atmosphere. The first of these is the claimed detection of water absorption in transmission made by \cite{2010MNRAS.409..963B}, based on the larger values for $\RpRs$ that those authors measured for the $5.8\um$ and $8.0\um$ channels relative to the $3.6\um$ and $4.5\um$ channels. The second is the inference of a thermal inversion in the atmosphere, based on the measurement of deeper eclipses in the $4.5\um$ and $5.8\um$ channels relative to the $3.6\um$ channel by \cite{2008ApJ...673..526K}.

Figure \ref{fig:transmission_and_emission_spectra} shows the transmission and emission measurements made in the current study with overplotted model spectra computed using the 1D radiative transfer code \texttt{ATMO} \citep{2014A&A...564A..59A, 2015arXiv150403334T}, assuming solar abundances and radiative-convective equilibrium. Models were generated under two different scenarios for the day-to-night heat redistribution efficiency:\ namely, uniform redistribution and zero redistribution (respectively, $\omega=0.5$ and $\omega=1$ in Figure \ref{fig:transmission_and_emission_spectra}). Number fractions as a function of atmospheric pressure are given for the major molecules in the lower left panel of Figure \ref{fig:transmission_and_emission_spectra} and the corresponding pressure-temperature (PT) profiles are shown in the lower right panel of Figure \ref{fig:transmission_and_emission_spectra}.

Our transmission results are in poor agreement with the clear-atmosphere model predictions. In particular, we do not see any enhancement in the opacity at $4.5\um$ relative to $3.6\um$ due to water and carbon monoxide absorption. As was mentioned in Section \ref{sec:results_rprs}, our transmission results are also in conflict with those originally presented by \cite{2010MNRAS.409..963B}. Specifically, Beaulieu et al.\ measured larger effective radii for the planet in the $5.8\,\um$ and $8.0\,\um$ channels, which coincide with a water absorption band, relative to the $3.6\,\um$ and $4.5\,\um$ channels. We, however, obtain significantly larger uncertainties in the $5.8\um$ channel and find that the effective radius is constant, or possibly decreasing modestly, across the $3.6$--$8.0\um$ IRAC wavelength range. The difference between the results obtained in the current study and those of Beaulieu et al.\ are likely due to the two approaches used for treating the ramp systematics. For the $5.8\,\um$ channel, Beaulieu et al.\ truncated the first section of the lightcurve and fit a linear trend in time to the remainder, and for the $8.0\,\um$ channel they decorrelated the ramp using a quadratic polynomial in logarithmic time. The GP model adopted in the current study should be capable of replicating both these explicit functional forms, and indeed, allows marginalisation over an even broader function space (Section \ref{sec:gpmodels}). We also confirmed that consistent results were obtained with the GP model when sections of varying duration were truncated from the start of the lightcurve. Based on our revised estimates of the planet's effective radii, we conclude that there is no evidence for water absorption in transmission over the IRAC bandpasses. Indeed, \cite{2013ApJ...774...95D} have measured a muted water feature at $1.4\um$ using WFC3, and interpreted this as evidence for haze in the atmosphere of HD\,209458b. Our results do not contradict this picture, and could suggest that the effect of the opacity inferred by Deming et al.\ at $1.4\um$ remains significant out to the IRAC wavelengths. 

We also modified the clear-atmosphere transmission model by fitting for an opaque cloud deck to simulate a grey opacity source across the IRAC wavelength range. This was done by allowing the clear atmosphere model to shift vertically while simultaneously setting the absorption to be constant at the cloud deck altitude. Thus our simple model had two tunable parameters:\ the overall vertical shift of the clear-atmosphere model and the cloud deck altitude. Our best-fit to the data gives a reduced $\chi^2$ of $1.4$, with an opaque cloud deck at $\RpRs = 0.1210 \pm 0.003$. As can be seen in the top left panel of Figure \ref{fig:transmission_and_emission_spectra}, the resulting ``cloudy'' model is simply a horiztonal line, implying that the IRAC data show no evidence for even reduced-amplitude absorption features extending above a cloud deck. If we instead fit a simple flat-line model to the data, where the only free parameter is the vertical level of the opaque cloud deck, the number of degrees of freedom increases from six to seven, and the reduced $\chi^2$ improves to $1.2$. Due to the size of our uncertainties, however, we cannot rule out muted absorption features of similar amplitude to the H$_2$O feature measured by \cite{2013ApJ...774...95D} at $1.4\um$ with WFC3. Furthermore, we note that our result is consistent at the $\sim 1\sigma$ level with the cloud deck altitude implied by the WFC3 transmission spectrum. Such a comparison should be treated with caution, however, as Deming et al.\ fixed $\aRs=8.95$ and $i=86.93^\circ$ in their lightcurve fits \citep[taken from][]{2007ApJ...655..564K}, which would introduce an offset in the absolute level those authors derive for $\RpRs$ relative to our study.  

For the emission, we produced models both with and without inverted pressure-temperature profiles (lower right panel of Figure \ref{fig:transmission_and_emission_spectra}). To produce thermal inversions in the former models, we artificially included titanium oxide (TiO) and vanadium oxide (VO) with a constant abundance throughout the atmosphere (lower left panel of Figure \ref{fig:transmission_and_emission_spectra}). Our emission results reinforce those of \cite{2014ApJ...796...66D}, who argued that there is no evidence for a thermal inversion in the atmosphere of HD\,209458b based on the revised estimates for the $4.5\um$ and $5.8\um$ eclipse depths. This can be seen in Figure \ref{fig:transmission_and_emission_spectra}, where the models with a thermal inversion are shown to provide a very poor match to the data. For the models without a thermal inversion, the match is better, but still poor. In particular, the latter models underpredict the emission in the $4.5\um$ channel, where an absorption feature due to carbon monoxide is expected to block radiation emitted from the planetary atmosphere. The fact that we do not detect this absorption is in line with the recent non-detection of carbon monoxide in the dayside hemisphere made by \cite{2015A&A...576A.111S} using high-resolution spectroscopy. 

We also fit the emission data with a model that assumes the planet radiates as an isothermal blackbody. To generate this model, an ATLAS spectrum \citep{1979ApJS...40....1K, 1993KurCD..13.....K} computed specifically for HD\,209458 was used for the stellar emission\footnote{http://kurucz.harvard.edu/stars/hd209458} and the radius ratio was fixed to $\RpRs = 0.121$. Assuming a Planck spectrum for the planet, a temperature of $1484 \pm 18$\,K was found to give the best fit to the data, with a reduced $\chi^2$ of 1.5. Indeed, \cite{2014MNRAS.444.3632H} have recently put forward the case that blackbody radiation can explain the majority of IRAC emission data that has been published for exoplanets to date, largely due to underestimated uncertainties for the eclipse depths. The acceptable fit provided by the isothermal blackbody for our data supports this hypothesis in the case of HD\,209458b, especially considering the reduced $\chi^2$ could be even lower if we marginalised over the covariance parameters and obtained larger uncertainties (Section \ref{sec:lcfitting}).

One possibility is that the emission measurements are probing an isothermal layer of the atmosphere above a cloud deck that extends across the dayside hemisphere. This could simultaneously explain the lack of absorption features detected in the IRAC transmission data, and the muted H$_2$O feature detected in the WFC3 transmission data, discussed above. We stress, however, that this is a speculative scenario, poorly constrained by the existing observations.

\section{Conclusion} \label{sec:conclusion}

We have presented an analysis of IRAC transit and eclipse lightcurves for HD\,209458b. By binning the lightcurves in time, it was possible to perform the lightcurve analyses using GP models. GPs allow transit lightcurve models to be elegantly defined within a rigorous Bayesian framework. They provide a natural mechanism for handling poorly-understood systematics in the data that are unrelated to the astrophysical signal of interest. Uncertainty is propagated through all levels of the model in a clear and transparent manner, and Occam's razor is automatically implemented, mitigating against overfitting. 

We have made a number of significant revisions to previously published results, and in many cases the uncertainties for inferred planet parameters have been increased by factors of $\sim$1--4. This can largely be attributed to the flexibility of the GP models, which allow complex correlations to be handled with a small number of free parameters. The latter point is important, as it means that marginalisation over the model parameter space remains computationally tractable, allowing uncertainties that realistically quantify the degeneracies between the planet signal and instrumental systematics to be derived. 

We obtain an overall improvement in consistency for the normalised semimajor axes $\aRs$ and orbital inclinations $i$ across different epochs and wavelength channels, compared to results that have been previously published for these datasets. This provides evidence that the GP models are effectively accounting for the systematics, and typically less prone to underestimating uncertainties compared with other lightcurve fitting approaches used in the literature. 

The revised GP analyses presented here draw into question a number of claims that have previously been made regarding the atmosphere of HD\,209458b, including the detection of water absorption in transmission and the inference of an inverted PT profile for the dayside hemisphere. Instead, our transmission measurements are consistent with a featureless spectrum, and our emission measurements are fit reasonably well assuming the planet radiates as an isothermal blackbody with a temperature of $1484 \pm 18$\,K.

Taken together, our results illustrate how sensitive IRAC analyses are to the systematics treatment. GP analyses have been shown to produce results that are generally more stable, and with uncertainties that are relatively conservative, compared to those obtained using other common approaches. However, we do tend to find good agreement with results obtained using pixel-mapping techniques in the $3.6\um$ and $4.5\um$ channels, which is unsurprising given that GPs are in essence quite similar to pixel-mapping. Nonetheless, many results have been published using simpler $xy$ polynomial decorrelations and parametric ramp models, which may not always be adequate. Combined with the lack of spectral resolution afforded by the broad bandpasses, our results suggest that statements made previously in the literature about exoplanet atmospheres relying heavily on the interpretation of IRAC data should be regarded with caution.

\section*{Acknowledgments}

The authors are grateful to Robert Zellem, Fr\'{e}d\'{e}ric Pont and David Sing for useful discussions. This work is partly supported by the European Research Council under the European Community's Seventh Framework Programme (FP7/20072013 Grant Agreement No.\ 247060). 

\bibliographystyle{apj}
\bibliography{hd209}
%\bibliography{general}

\appendix

\clearpage
\section[]{Covariance parameter maximum likelihood estimates}

Table \ref{table:covpar_mle} reports the maximum likelihood estimates for the GP covariance parameters that were fixed for the MCMC analyses, as described in Section \ref{sec:lcfitting}.

\begin{table}
\begin{minipage}{\linewidth} 
\centering
\caption{Maximum likelihood estimates for the GP covariance parameters. These are the values that were fixed for the MCMC analyses described in Section \ref{sec:lcfitting}. \label{table:covpar_mle}}
\begin{tabular}{ccccccccc}
\hline \\
Channel & Date & Signal & $A_{xy}$ & $L_x$ & $L_y$ & $A_{t}$ & $L_{t}$ & $\sigma_w$  \\
($\mu$m) & & & (\%) & (pix) & (pix) & (\%) & (min) & (ppm) \smallskip \\
\hline \\

3.6 & 2005 Nov 28 & Eclipse &  $2.0552$ & $0.642$ & $0.528$ & $0.0444$ & $37.117$ & $1541$ \smallskip \\ 
    & 2007 Dec 31 & Transit &  $0.4874$ & $0.169$ & $0.187$ & $0.0216$ & $20.708$ & $2185$ \smallskip \\ 
    & 2008 Jul 19 & Transit &  $0.3089$ & $0.138$ & $0.136$ & $0.0677$ & $0.319$ & $2066$ \smallskip \\ 
    & 2011 Jan 12 & Eclipse (1st) &  $1.0988$ & $0.176$ & $0.091$ & $0.0261$ & $5.471$ & $499$ \smallskip \\ 
    & 2011 Jan 14 & Transit &  $0.6342$ & $0.127$ & $0.145$ & $0.1754$ & $0.322$ & $344$ \smallskip \\ 
    & 2011 Jan 16 & Eclipse (2nd) &  $1.8569$ & $0.235$ & $0.233$ & $0.0331$ & $6.710$ & $541$ \smallskip \\ 
    & 2014 Feb 13 & Eclipse (1st) &  $1.5366$ & $0.167$ & $0.135$ & $0.0398$ & $5.574$ & $553$ \smallskip \\ 
    & 2014 Feb 15 & Transit &  $1.6826$ & $0.195$ & $0.196$ & $0.0488$ & $0.553$ & $660$ \smallskip \\ 
    & 2014 Feb 17 & Eclipse (2nd) &  $1.2590$ & $0.151$ & $0.118$ & $0.0234$ & $3.089$ & $453$ \medskip \\ 
4.5 & 2005 Nov 28 & Eclipse &  $0.8656$ & $0.238$ & $0.254$ & $0.8796$ & $2386.526$ & $1391$ \smallskip \\ 
    & 2008 Jul 22 & Transit &  $0.0001$ & $143.003$ & $109.329$ & $0.0302$ & $5.887$ & $3148$ \smallskip \\ 
    & 2010 Jan 18 & Eclipse (1st) &  $0.2467$ & $0.067$ & $0.090$ & $0.0211$ & $25.212$ & $578$ \smallskip \\ 
    & 2010 Jan 19 & Transit &  $0.8769$ & $0.255$ & $0.231$ & $0.0086$ & $24.042$ & $569$ \smallskip \\ 
    & 2010 Jan 21 & Eclipse (2nd) &  $0.8468$ & $0.235$ & $0.270$ & $0.0182$ & $15.931$ & $566$ \medskip \\

\hline \\

Channel & Date & Signal & $A_{\tau}$ & $L_{\tau}$ & $h$ & $A_{t}$ & $L_{t}$ & $\sigma_w$   \\
($\mu$m) & & & (\%) & ($\log[\tn{min}]$) & (min) & (\%) & (min) & (ppm)  \smallskip \\
\hline \\

5.8 & 2005 Nov 28 & Eclipse &  0.0732 & $6.187$ & $2796.4$ & $0.0575$ & $195.8$ & $3956$ \smallskip \\ 
   & 2007 Dec 31 & Transit &  0.3080 & $6.127$ & $489.8$ & $0.0000$ & $67.9$ & $3037$ \smallskip \\ 
   & 2008 Jul 19 & Transit &  0.7993 & $12.986$ & $0.1$ & $0.0000$ & $>10^4$ & $2986$ \medskip \\ 
8.0 & 2005 Nov 28 & Eclipse &  0.2781 & $6.277$ & $1142.0$ & $0.0000$ & $>10^4$ & $3749$ \smallskip \\ 
   & 2007 Dec 24 & Transit &  0.0000 & $357.352$ & $291853.4$ & $0.2427$ & $1221.2$ & $1163$ \smallskip \\ 
   & 2007 Dec 25 & Eclipse &  0.0163 & $5.887$ & $12729.4$ & $0.0000$ & $>10^4$ & $1215$ \smallskip \\ 
   & 2008 Jul 22 & Transit &  1.8164 & $9.791$ & $1.0$ & $0.0000$ & $>10^4$ & $2213$ \\ \\ \hline

\end{tabular}
\end{minipage}
\end{table}

\bsp
\label{lastpage}
\end{document}